\documentclass[pra,twocolumn,superscriptaddress,showpacs,amsmath,amssymb]{revtex4-2}

\usepackage{graphicx}
\usepackage{dcolumn}
\usepackage{bm}
\usepackage{epsfig}
\usepackage{amsmath}
\usepackage{amssymb}
\usepackage{color}
\usepackage{bbm}
\usepackage{CJK}
\usepackage[colorlinks=true,citecolor=blue,linkcolor=blue,urlcolor=blue]{hyperref}
\begin{document}

\title{Switchable Dissipative Ising coupling Based on Three-Body Coupling in magnon systems}
\author{Xi-Wen Dou}
\affiliation{Zhejiang Key Laboratory of Quantum State Control and Optical Field Manipulation, Department of Physics, Zhejiang Sci-Tech University, 310018 Hangzhou, China}
\author{Zheng-Yang Zhou}
\altaffiliation[zheng-yang.zhou@zstu.edu.cn]{}
\affiliation{Zhejiang Key Laboratory of Quantum State Control and Optical Field Manipulation, Department of Physics, Zhejiang Sci-Tech University, 310018 Hangzhou, China}
\author{Ai-Xi Chen}
\altaffiliation[aixichen@zstu.edu.cn]{}
\affiliation{Zhejiang Key Laboratory of Quantum State Control and Optical Field Manipulation, Department of Physics, Zhejiang Sci-Tech University, 310018 Hangzhou, China}
\date{\today}

\begin{abstract}
Magnonic systems present a compelling platform for quantum technology, owing to their strong capacity to form hybrid quantum systems via diverse couplings. To unlock the full potential of these systems, the engineering of flexible coupling between multiple magnon modes is essential. Here, we propose a method to realize switchable dissipative Ising coupling in magnon systems, leveraging the three-body coupling among photon, phonon, and magnon. This type of dissipative coupling is a critical component for constructing Ising machines designed to solve complex combinatorial optimization problems. By dynamically tuning the phase of a nonlinear mechanical pump, we demonstrate the realization of both ferromagnetic and antiferromagnetic dissipative interactions. The validity of the scheme is confirmed by numerical simulations, which also demonstrate its robustness against a strong uncontrollable part of dissipation. Our work provides a versatile tool that can facilitate the implementation of magnon-based quantum computing and the exploration of many-body magnon physics.
\end{abstract}
\maketitle


%
%


\section{Introduction}                           
Magnons, the quanta of spin-wave excitations in magnetically ordered materials~\cite{1magnonmaterial1,1magnonmaterial2,1magnonmaterial3,1magnonmaterial4}, have recently emerged as a versatile platform for quantum optics, owing to their intrinsic nonlinear characteristics~\cite{2magnonnonlinearity1,2magnonnonlinearity2,2magnonnonlinearity3}, tunable scattering behavior~\cite{3magnonscattering1,3magnonscattering2,3magnonscattering3}, and potential applications in spintronics~\cite{4magnonspintronics1,4magnonspintronics2,4magnonspintronics3}.\ These characteristics enable strong hybrid coupling between magnons and other subsystems, leading to a wide range of applications in hybrid quantum architectures~\cite{5hybrid-magnon-system1,5hybrid-magnon-system2,5hybrid-magnon-system3,5hybrid-magnon-system4}.\ Notably, the intrinsic Kerr nonlinearity~\cite{8kerr-decoherence2} of magnons provides a powerful resource for generating nonclassical states, such as Schrödinger cat states~\cite{6magnon-kerr-cat1,6magnon-kerr-cat2,6magnon-kerr-cat3,6magnon-kerr-cat4} and squeezed states~\cite{7magnon-kerr-squeeze1,7magnon-kerr-squeeze2,7magnon-kerr-squeeze3}.\ Among various magnonic systems, yttrium iron garnet (YIG) microspheres are distinguished by their exceptionally high quality factors and strong Kerr nonlinearity~\cite{7magnon-kerr-squeeze3,9YIG-kerr2,9YIG-kerr3}, making them ideal platforms for cavity quantum electrodynamics experiments~\cite{10YIG-in-QED1,10YIG-in-QED2}.\ An intriguing phenomenon in magnonic systems is bistability~\cite{11magnon-bistable1,6magnon-kerr-cat3,11magnon-bistable3,11magnon-bistable4,11magnon-bistable5,11magnon-bistable6}, which can serve as an analog for simulating Ising spins~\cite{12-scattering-magnon-bistable}.

Ising spins can be used to solve combinatorial optimization problems~\cite{1CIMusedtosolvingNPhard1,1CIMusedtosolvingNPhard2,1CIMusedtosolvingNPhard3}, which can describe many practical problems~\cite{10.1021/acs.jctc.3c00943,10.1007/s11433-023-2147-3} and are known to be NP-hard. To efficiently tackle such combinatorial optimization problems, coherent Ising machines (CIMs)~\cite{2CIMusedtogeneratingneuralnetwork1,2CIMusedtogeneratingneuralnetwork2,3CIMusedtomakinglogiccircuits1,3CIMusedtomakinglogiccircuits2,3CIMusedtomakinglogiccircuits3} have been developed, which exploit optical bistability to represent and manipulate Ising spins.\ CIMs have demonstrated excellent scalability and robustness against noise in both theoretical analyses and experimental implementations~\cite{4CIMinexperiment1,4CIMinexperiment2,4CIMinexperiment3}. In addition to their original implementation based on optical parametric oscillators and fiber networks, CIMs have also been realized on various alternative platforms~\cite{2CIMusedtogeneratingneuralnetwork2,5CIMbasedon3,5CIMbasedon5}.\ Recent studies indicate that bistable states in magnon systems can potentially be utilized to construct CIMs~\cite{12-scattering-magnon-bistable}. In addition to the implementation of effective Ising spins, the realization of effective couplings is essential for CIMs.\ Moreover, these couplings should be reconfigurable to simulate arbitrary Ising models.\ In fiber-based CIMs, the switching of coupling terms is usually achieved by electro-optic modulators~\cite{10.1038/nphoton.2014.249} or field-programmable gate arrays~\cite{10.1126/science.aah5178,1CIMusedtosolvingNPhard3}. The design of switchable Ising couplings in magnon systems, however, remains an open challenge.

Three-body couplings in magnon systems~\cite{santiouhe} provide a potential mechanism for realizing switchable Ising couplings in magnon-based CIMs. In particular, magnon systems, which couple photons, phonons, and magnons, provide an ideal platform for exploring three-body interactions~\cite{3photons-phonons-and-magnons-1,3photons-phonons-and-magnons-2,3photons-phonons-and-magnons-3,3photons-phonons-and-magnons-4,3photons-phonons-and-magnons-5}. These hybrid magnon systems have attracted considerable attention in recent years owing to their ability to enhance both coherent couplings~\cite{9YIG-kerr3,1Multi-enhanced-coupling2,1Multi-enhanced-coupling3,9YIG-kerr3} and dissipative couplings~\cite{2Multi-dissipative-coupling1,2Multi-dissipative-coupling2}. The interactions in these systems can be tuned via either squeezing~\cite{4triple-squeeze-1,4triple-squeeze-2,6partite-control-partite1,6partite-control-partite2,6partite-control-partite3} or coherent feedback control~\cite{5triple-feedback-control1,5triple-feedback-control2}.In addition, the manipulation of steady states~\cite{7dissipation-with-stable-states1,7dissipation-with-stable-states2,7dissipation-with-stable-states3,7dissipation-with-stable-states4} through dissipation has been extensively studied in these systems, providing guidance for the design of Ising couplings in magnon-based CIMs.

In this paper, we propose constructing switchable dissipative couplings in magnon-based CIMs via three-body interactions among phonons, photons, and magnons. The photon mode induces loss in the magnon mode, while the phonon mode modulates the loss intensity. We further show that the same mechanism can be applied to dissipative couplings, enabling a transition from ferromagnetic to antiferromagnetic interactions. The performance of the scheme is subsequently evaluated through numerical simulations. Our numerical results demonstrate that, in a two-magnon system with switchable dissipative coupling, the steady states can exhibit either in-phase or anti-phase behavior depending on the mechanical state. Moreover, the switchable coupling can also modify the relative phases of the transient states. In addition to the ideal case, the effect of unwanted losses is also evaluated to demonstrate the robustness of the method.

This paper is organized as follows. In Sec.~\ref{single-mode}, we present the basic concept of designing a switchable loss channel and verify it numerically.
In Sec.~\ref{COLLECTIVE DISSIPATION}, we extend this concept to dissipative coupling and demonstrate that channel switching enables rapid state transitions.
We characterize these transitions using expectation values and dual-mode Wigner functions. Section~\ref{conclusion} is the conclusion.

\section{Basic concepts for switchable loss}
\label{single-mode}
Previous work~\cite{12-scattering-magnon-bistable} has shown that bistable states in magnon modes can be used to emulate spin-down and spin-up configurations. To achieve this, the magnon mode is first parametrically pumped by the Hamiltonian:
\begin{equation}
\label{squeezing}
H_{\rm s} = s_a^{*}{a^\dagger}{a^\dagger}+ {s_a}aa,
\end{equation}
where $a$ is the annihilation operator of the magnon mode, and $s_a$ represents the effective nonlinear pumping intensity.\ To construct bistable states, a self-Kerr nonlinearity is introduced in the magnon mode,
\begin{equation}
\label{self-kerr}
H_{\text{self-Kerr}} = K{a^\dagger}{a^\dagger}aa,
\end{equation}
where $K$ is the self-Kerr coefficient. This fourth-order nonlinearity induces a shift in the magnon frequency and facilitates the emergence of bistability, as detailed in Ref.~\cite{self-kerr}. Furthermore, single-photon loss also plays a crucial role. Such a dissipative effect is typically modeled by a Lindblad term in the master equation,
\begin{eqnarray}
\label{master-equation}
\frac{\partial}{\partial t}\rho&=&i[H,\rho]+\frac{\gamma_{\rm s}}{2}\mathcal{L}(\rho,a),\nonumber\\
\mathcal{L}(\rho,L)&\equiv&(2L\rho L^{\dag}-L^{\dag}L\rho-\rho L^{\dag}L),
\end{eqnarray}
with the Lindblad operator $L = a$ and the loss rate $\gamma_{\rm s}$.

The emergence of bistable magnon states is contingent upon the squeezing parameter $s_a$, the loss rate $\gamma_s$, and the Kerr coefficient $K$ residing within appropriate regimes. The interplay between the self-Kerr nonlinearity and the mechanical mode squeezing governs the system's steady-state behavior, where these states can be stable or unstable contingent upon the magnitude of the nonlinearity and squeezing, as noted in Ref.~\cite{12-scattering-magnon-bistable}.

The bistable magnon states can simulate spin states and the corresponding Ising model,
\begin{equation}
H = -\sum_{i,j} J_{i,j} \sigma_i \sigma_j,
\end{equation}
where $\sigma_{i} = \{-1,1\}$ and  $J_{i,j}$  denote  the  value  of  the  $i_{th}$  spin  and  a  coupling coefficient between the $i_{th}$ and $j_{th}$ spins, respectively. Following the approach outlined in Ref.~\cite{12-scattering-magnon-bistable}, the effective spins can be constructed as,
\begin{eqnarray}
    \vert \alpha \rangle &\leftrightarrow& \vert \uparrow \rangle,\nonumber\\
    \vert -\alpha \rangle &\leftrightarrow& \vert \downarrow \rangle.
\end{eqnarray}
Distinct Ising coupling terms are realized through varying collective dissipation terms. The coupling term between the $i$th spin and the $j$th spin $J_{i,j}\sigma_z^i\sigma_z^j$ corresponds to a collective loss term between the $i$th and the $j$th magnon mode. For $J_{i,j}>0$, the two spins tend to adopt opposite orientations, consequently lowering the system's energy. Under such circumstances, a dissipation term,
\begin{eqnarray}
\label{a1+a2}
\frac{\Gamma_{\rm c}}{2}\mathcal{L}(\rho,a_i+a_j),
\end{eqnarray}
is incorporated into the Ising machines, thereby inducing the two magnon modes to favor opposite phases. Conversely, a negative $J_{i,j}$ corresponds to a distinct Lindblad term,
\begin{eqnarray}
\label{a1-a2}
\frac{\Gamma_{\rm c}}{2}\mathcal{L}(\rho,a_i-a_j).
\end{eqnarray}
Consequently, the collective dissipation terms must be switchable to ensure the convenient handling of diverse Ising problems by the machine.

\begin{figure}[t]
\center 
\includegraphics[width=3in]{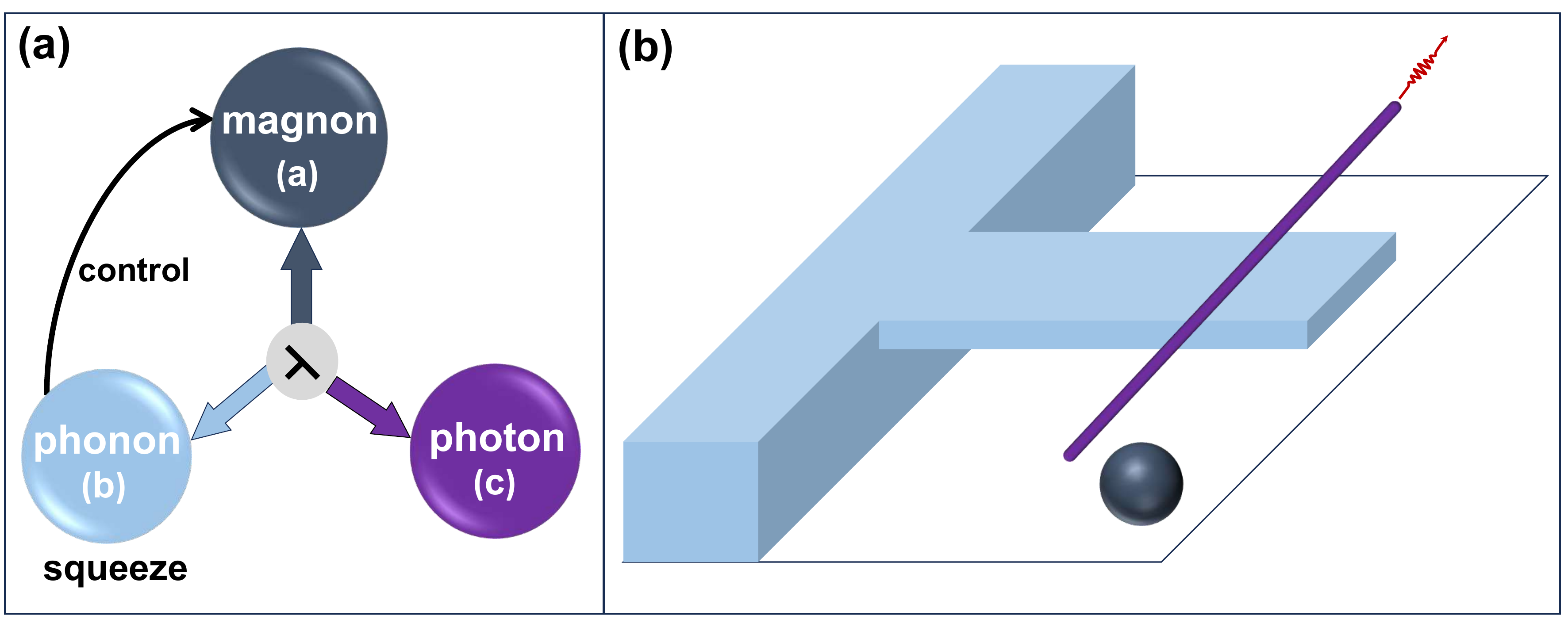} 
\caption{(a)~Schematic diagram of the triple-body system. The system involves three distinct modes: the magnon mode ($a$), the phonon mode ($b$), and the photon mode ($c$).\ The operators $a^\dagger$, $b^\dagger$, and $c^\dagger$ denote the creation operators for the magnon, phonon, and photon modes, respectively.\ $\lambda$ represents the coupling strength among the three modes. (b)~Physical realization of the hybrid system.\ The system is composed of a magnon mode (a YIG microsphere), a phonon mode (a mechanical cantilever), and a photon mode (a transmission line). The YIG sphere is placed below the cantilever.\ The transmission line (indicated by the purple line) is positioned above the cantilever, illustrating the geometry of the coupling elements.} 
\label{fig1} 
\end{figure}

A viable approach for achieving switchable dissipation channels is the manipulation of dissipation via three-body coupling within magnon systems \cite{santiouhe}. As illustrated in Fig.~\ref{fig1}, we consider a system comprising a magnon mode (a YIG microsphere), a photon mode (a transmission line), and a phonon mode (a cantilever). The interaction Hamiltonian is given by:
\begin{equation}
\label{santi-term}
H_{\text{int}} = \lambda (b + {b^\dagger})(a{c^\dagger} + {a^\dagger}c),
\end{equation}
where ${a}$, ${b}$, and ${c}$ denote the annihilation operators for the magnon, phonon, and photon modes, respectively. Following the procedure of adiabatic elimination, this interaction Hamiltonian is capable of inducing a dissipation channel:
\begin{equation}
\label{single-mode-conloss}
L = -(b + {b^\dagger})a,
\end{equation}
this channel is associated with a dissipation rate $\Gamma_{\rm control}$ (see also Appendix A). By manipulating the quantum state of the phonon mode, the effective loss rate in the magnon mode can be precisely tuned.

Reference~\cite{santiouhe} demonstrated that the coupling strength between the magnon, spin, and phonon modes can be enhanced by applying squeezing to the mechanical mode:
\begin{equation}
U(r) = \exp \left( \frac{r ({b}^2 - {b}^{\dagger 2})}{2} \right),
\end{equation}
where $r$ is the squeezing parameter. This method further allows for the potential tuning of the loss rate of the dissipation channel induced by the three-body interaction in Eq.~(\ref{single-mode-conloss}). Given that the Lindblad operator in Eq.~(\ref{single-mode-conloss}) is proportional to the dimensionless position operator of the mechanical mode, $x\equiv(b+{b}^{\dagger})/\sqrt{2}$, the squeezing or anti-squeezing of the position may modulate the decay rate.

It is noteworthy that the objective of this section is to modulate the overall system's net dissipation rate using the squeezing effect. We will first introduce the theoretical model for mechanical mode squeezing incorporating loss and nonlinear pump. Subsequently, the control effects of mechanical-mode squeezing on magnon-mode dissipation will be elucidated. While the detailed design for switchable collective loss will be presented in the subsequent section.

\subsection{Effects of Mechanical-Mode Squeezing}
To elucidate how phonon squeezing in the dissipation channel characterized by Eq.~(\ref{single-mode-conloss}) affects the loss of the magnon mode, we initially consider the following squeezing terms in the mechanical mode \cite{squeeze--pump},
\begin{equation}
    H_{\rm b} = s_b^* b^\dagger{b}^\dagger, + s_b {b}{b}
\label{phonon_squeeze}
\end{equation}
where $b$ is the phonon annihilation operator, and $s_b$ is the squeezing parameter, which governs the degree of squeezing in either the displacement or momentum quadrature. It should be noted that we employ the displacement and momentum operators of the phonon mode here,

\begin{equation}\label{dimensionlessquadrature}
x = \frac{{b}+{b}^\dagger}{\sqrt{2}}, \quad p = i\frac{{b}^\dagger-{b}}{\sqrt{2}}. 
\end{equation}

The squeezed quadrature is determined by the phase of $s_b$, namely, the phase of the pump field. A real positive $s_b$ induces position squeezing ($\langle(x-\langle x\rangle)^2\rangle<\frac{1}{2}$), whereas a real negative $s_b$ yields momentum squeezing ($\langle(p-\langle p\rangle)^2\rangle<\frac{1}{2}$)\cite{POSI2NEGA}.

To maintain the mechanical mode in its steady state, we introduce a strong dissipation term described by the Lindblad operator:
\begin{equation}\label{lossofb}
    L_{b} = b, 
\end{equation}
with a dissipation rate $\Gamma_b = 10s_a$. Such a strong dissipation manifests a two-fold effect: Firstly, it suppresses the mechanical mode from entering the parametric instability regime. Secondly, this dissipation term suppresses the back-action of the magnon mode on the mechanical mode during the process of controlling the magnon mode via the mechanical mode.

\begin{figure}[t] 
\center 
\includegraphics[width=3.4in]{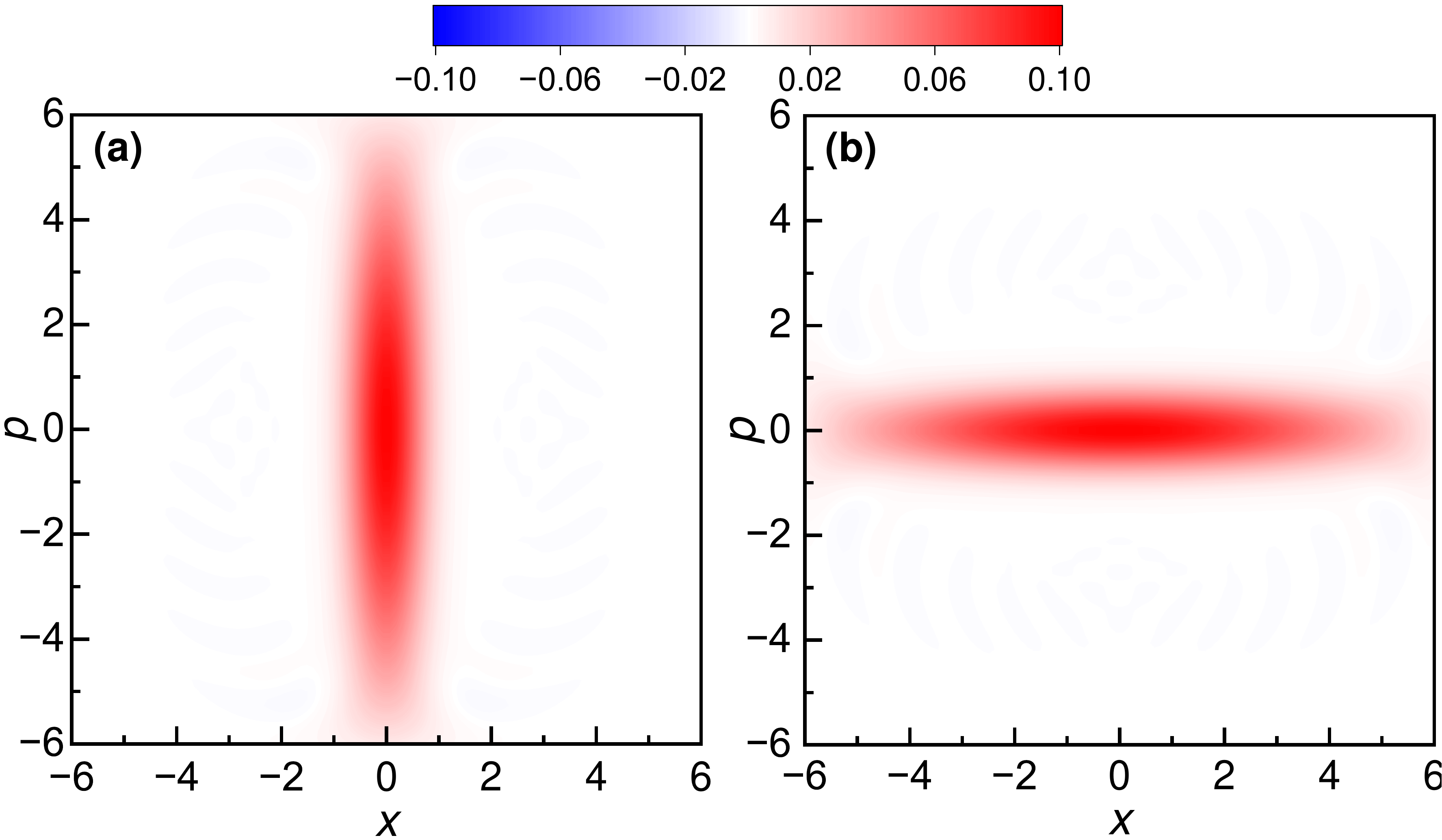} 
\caption{The wigner function of the squeezed phonon mode in the phase space. The Fock-space truncation is $N=25$. The evolution time is set to 20(in units of $s_b$). (a)~The state is squeezed in the displacement quadrature, corresponding to the effective squeezing parameter $s_b=2.4i$, (b)~The state is squeezed in the momentum quadrature, corresponding to $s_b=-2.4i$.} 
\label{fig2} 
\end{figure}

We perform numerical simulations of the mechanical system's evolution using the qutip package, and the resultant Wigner function after a sufficiently long evolution time is illustrated in Fig.~\ref{fig2}. Corresponding to different pump phases, squeezing is observed in the position quadrature and the momentum quadrature, respectively. It is straightforward to anticipate that such squeezing can influence the dissipation described in Eq.~(\ref{single-mode-conloss}).

\subsection{Switchable loss in single-magnon system}
This section investigates the influence of squeezing in the mechanical mode $b$ on the magnon mode $a$ and presents the foundational concept of switchable dissipation. The magnon mode under investigation is subject to a nonlinear pump~(\ref{squeezing}), a nonlinear Kerr term~(\ref{self-kerr}), and controllable dissipation~(\ref{single-mode-conloss}). Under the strong dissipation regime, the magnon steady state is characterized as a squeezed state, whereas bifurcation occurs in the weak dissipation limit \cite{10.1103/c91r-8t3h}. In order to modulate the dissipation strength of the magnon mode, we introduce a two-phonon pump term~(\ref{phonon_squeeze}) and a dissipation term~(\ref{lossofb}) for the mechanical mode. As demonstrated in Fig.~\ref{fig2}, the squeezed quadrature can be selected via the pump phase, so that the loss term shown in Eq.~(\ref{single-mode-conloss}) can be adjusted.

The Hamiltonian of the system studied is, 
\begin{equation}
H_{\text{total}} = s_a^{*}{a^\dagger}{a^\dagger}+ s_aaa+ K {a}^\dagger {a}^\dagger{a}{a}+ s_b^*{b}^\dagger{b}^\dagger + s_bbb .
\end{equation}
where $s_a$ and $K$ are, respectively, the nonlinear pump coefficient and the Kerr strength of the magnon mode. The nonlinear pump coefficient of the mechanical mode is described by $s_b$. The system's overall dissipation is described by two loss channels,
\begin{eqnarray}
L_{\rm control}&=&-(b + {b^\dagger})a,\nonumber\\
L_b&=&b.
\end{eqnarray}
Consistent with the preceding sections, we designate the loss rate of the mechanical mode as $\Gamma_b$ and that of the controllable channel as $\Gamma_{\rm control}$. 
\begin{figure}[t] 
\center 
\includegraphics[width=3in]{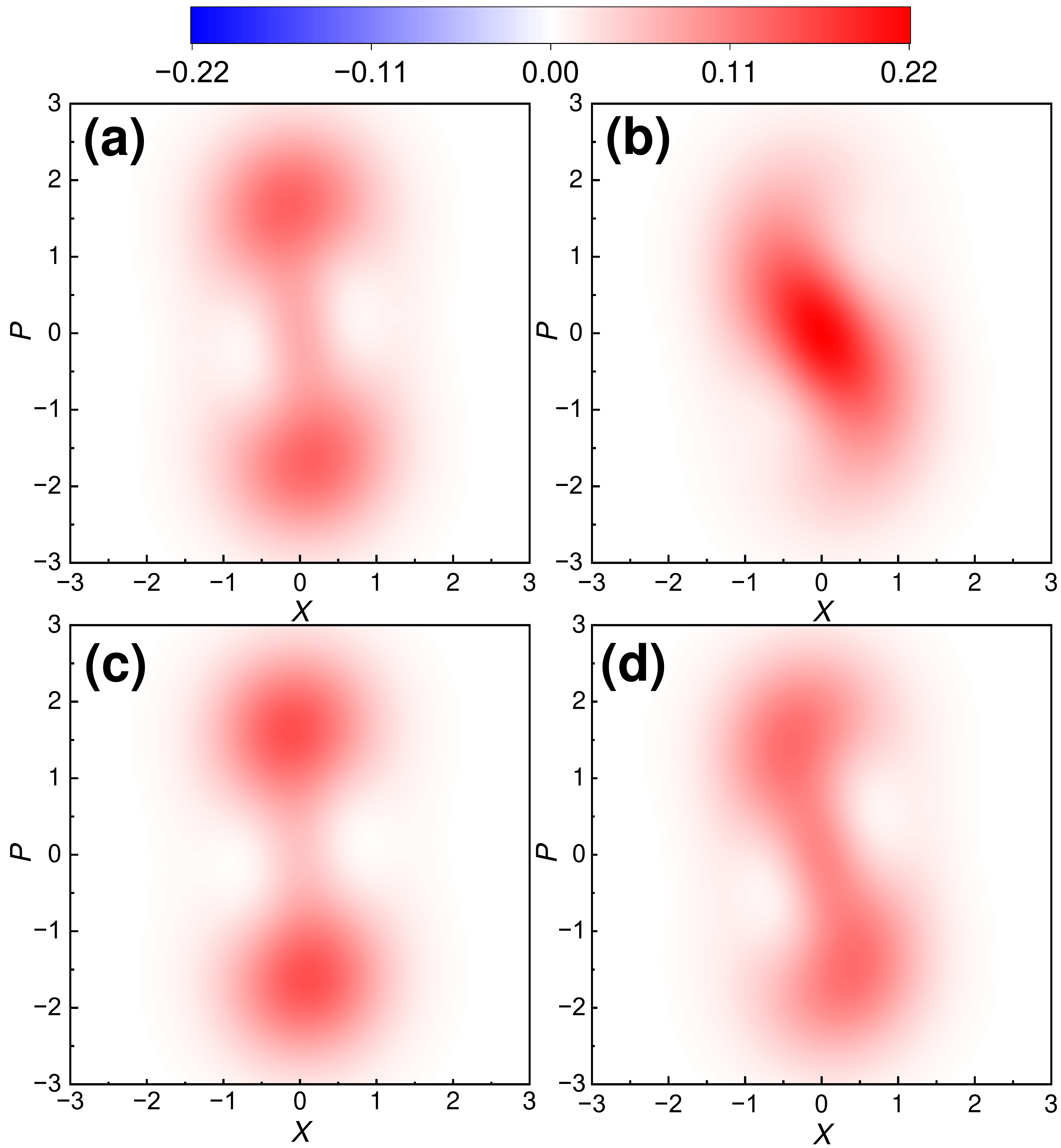} 
\caption{The Wigner function of the magnon mode ($a$) under the effects of phonon squeezing and intrinsic magnon dissipation. The Fock-space truncation is $N=20$. The coupling strength is $\lambda_s=0.8$, the Kerr non-linearity is $K=0.75$, and the driving amplitude is $s_a=1$. The evolution time is set to $t=20$ (in units of $s_a$).The subfigures illustrate the magnon Wigner function under different phonon squeezing parameters ($s_b$):(a) Strong squeezing in the displacement quadrature, $s_b = 2.4i$.(b) Strong squeezing in the momentum quadrature, $s_b = -2.4i$.(c) Moderate squeezing in the displacement quadrature, $s_b = 1.5i$.(d) Moderate squeezing in the momentum quadrature, $s_b = -1.5i$.}   
\label{fig3}
\end{figure}
We prepare the initial state of the system in the vacuum state and perform numerical simulation of its dynamical evolution. Fig.~\ref{fig3} illustrates the long-time-limit Wigner functions of the system under various parameter settings. The magnon squeezing coefficient is set to $s_a=1$, and is employed as the common unit for all other parameters. The truncation dimension of the Hilbert space is set to $N=20$. The remaining parameters are specified as $\Gamma_b=10$, $\Gamma_{\rm control}=0.64$, and $K=0.75$. In Figs.~\ref{fig3} (a) and (b), we adopt a strong nonlinear pump $|s_b|=2.4$, a value close to the parametric stability threshold. In Fig.\ref{fig3}(a), the phonon squeezing coefficient is set to the position quadrature $s_b = 2.4i$, and the Wigner function exhibits two separated components, unequivocally revealing the emergence of a bistable state. This bistable characteristic is indicative of a weak dissipation in the magnon mode. In Fig.\ref{fig3} (b), the mechanical mode squeezing is switched to the momentum quadrature, with $s_b=-2.4i$. The corresponding Wigner function transforms into a single-component squeezed state, which indicates the presence of strong dissipation in the magnon mode. In Figs.~\ref{fig3} (c) and (d), we consider a weaker squeezing intensity for the mechanical mode $|s_b|=1.5$, and demonstrate the performance of the switchable dissipation. In Fig.~\ref{fig3} (c), the squeezing is in the position quadrature, consequently leading to a clear separation of the two components in the Wigner function. In Fig.~\ref{fig3} (d), the squeezing is switched to the momentum quadrature; however, the Wigner function still exhibits two less-separated components. Therefore, even though the adjustable range is smaller, weak squeezing can still realize a tunable dissipation channel.

\section{Switchable collective loss with three-body coupling}
\label{COLLECTIVE DISSIPATION}
\begin{figure}[t]
\center
\includegraphics[width=3in]{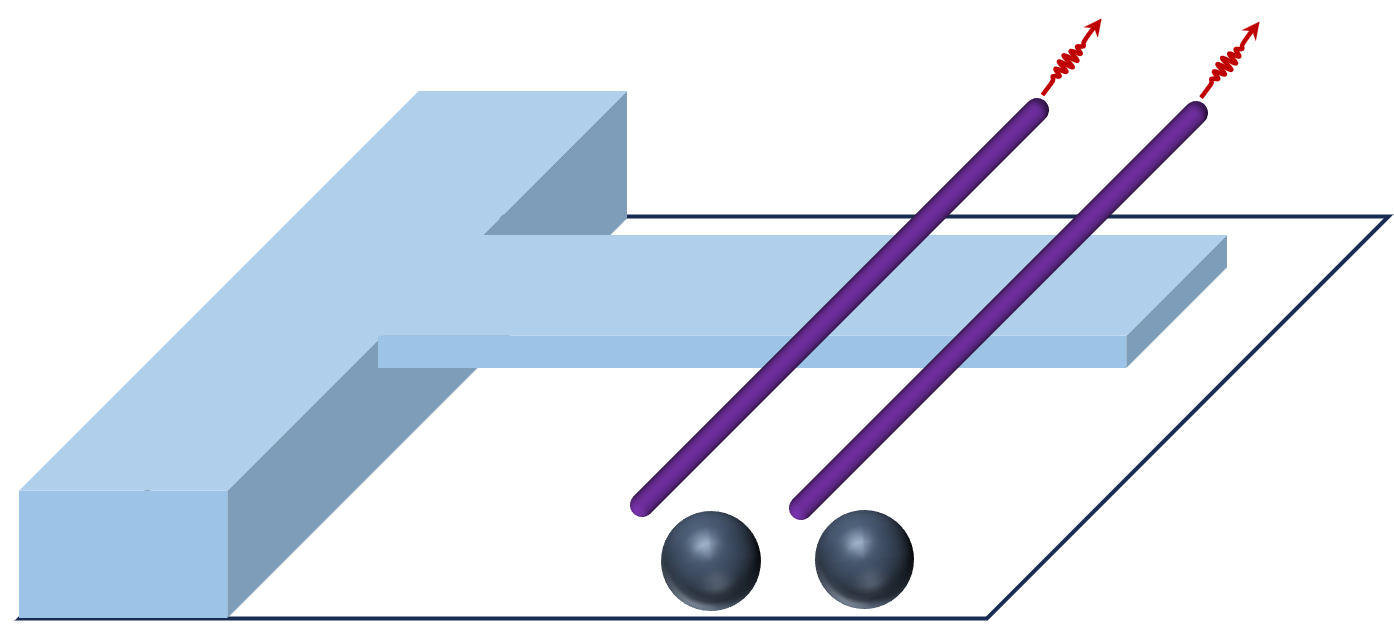}
\caption{Schematic illustration of an extended hybrid system module. This system incorporates two YIG microspheres placed below the mechanical cantilever. Two transmission lines (indicated in purple) are positioned above the cantilever.}
\label{fig4}
\end{figure}

In Sec.~\ref{single-mode}, we introduced the fundamental concept of controlling the loss in a magnon mode via three-body coupling and squeezing. We now extend this idea to collective loss and achieve switchable dissipative coupling. As illustrated in Fig.~\ref{fig4}, the system under consideration comprises two YIG microspheres, two transmission lines, and a single cantilever. Two conducting wires introduce distinct forms of collective dissipation to these two magnon modes, which are defined in Eqs.~(\ref{a1+a2}) and (\ref{a1-a2}). In addition, these two collective loss channels can be adjusted by the motion of the cantilever so that the dominant collective loss between the two magnon modes can be switched. 

In this system, the squeezing of the mechanical mode plays a pivotal role in regulating the collective dissipation. The Hamiltonian and the dissipation of this part remain identical to those presented in Eqs.~(\ref{phonon_squeeze}) and (\ref{lossofb}). Concurrently, nonlinear pump and Kerr terms are present in both YIG microspheres, which can be described by,
\begin{eqnarray}\label{two-magnon}
\begin{aligned}
    H_{\text{s}} =& H_{\text{s,1}}+H_{\text{s,2}}\\=&s_a^*{a_1}^\dagger{a}_1^\dagger+s_a{a}_1{a}_1+s_a^*{a}_2^\dagger{a}_2^\dagger+s_a{a}_2{a}_2,
\end{aligned}
\end{eqnarray}
and,
\begin{equation}\label{two-kerr}
\begin{aligned}
     H_{\text{self-kerr}} &=  H_{\text{self-kerr,1}}+ H_{\text{self-kerr,2}}\\&=K{a}_1^\dagger{a}_1^\dagger{a}_1{a}_1+K{a}_2^\dagger{a}_2^\dagger{a}_2{a}_2.
\end{aligned}
\end{equation}
Two conducting wires can be modeled as two cavity modes with strong loss, 
\begin{eqnarray}
\begin{aligned}
    H_c = &\omega_{c}{c_1}{^\dagger}{c_1}+\omega_{c}{c_2}{^\dagger}{c_2},\\
    &L_1=c_1,~~L_2=c_2,
\end{aligned}
\end{eqnarray}
with cavity frequency $\omega_c$ and cavity mode loss rate $\Gamma_{\rm cav}$. The three-body interactions among photon, phonon, and magnon modes are described by:
\begin{equation}\label{threebodyinteractions}
\begin{aligned}
    H_{\text{int,1}}=& \lambda_{\rm c}({b}+{b}^\dagger)[({a}_1+{a}_2){c_1}{^\dagger}+({a}_1+{a}_2)^\dagger{c_1}],\\
    H_{\text{int,2}}=& \lambda_{\rm c}i({b}-{b}^\dagger)[({a}_1-{a}_2){c_2}{^\dagger}+({a}_1-{a}_2)^\dagger{c_2}].
\end{aligned}
\end{equation}
Here, $\lambda_{\rm c}$ represents the coupling strength among the different components. These two cavity modes exhibit different coupling mechanisms to the magnon and mechanical modes, with the interactions defined by $H_{\text{int2}}$ and $H_{\text{int1}}$, respectively. Regarding the cavity-magnon coupling, one cavity mode couples in-phase to the two magnon modes, while the other cavity mode couples out-of-phase to the same two magnon modes. Such distinct collective couplings can induce the different collective dissipations presented in Eqs.~(\ref{a1+a2}) and (\ref{a1-a2}). To enable switching between the different dissipation channels, the two cavity modes also exhibit distinct coupling forms to the mechanical mode. These differing coupling forms can be adjusted by varying the phases (see also Appendix B).

After the adiabatic elimination of the two cavity modes, the interaction terms in Eq.~(\ref{threebodyinteractions}) turn into two distinct loss channels described by the Lindblad operators,
\begin{eqnarray}
\begin{aligned}\label{contralcollective}
        L_1 &= ({b}+{b}{^\dagger})({a}{_1}+{a}_2),\\
        L_2 &= i({b}{^\dagger}-b)({a}{_1}-{a}_2),
\end{aligned}
\end{eqnarray}
with the loss rate $\Gamma_{\rm cc}$. Although the loss rate of these nonlinear Lindblad operators is given by the nonlinear coupling strength $\lambda_{\rm c}$ and the cavity mode loss, the dissipation in the magnon modes can still be influenced by the state of the mechanical mode. Given that these two dissipation channels are related to the position and momentum quadratures, respectively, we can utilize squeezing in different quadratures to control the active dissipation channel.

\begin{figure}[t]
\center
\includegraphics[width=3in]{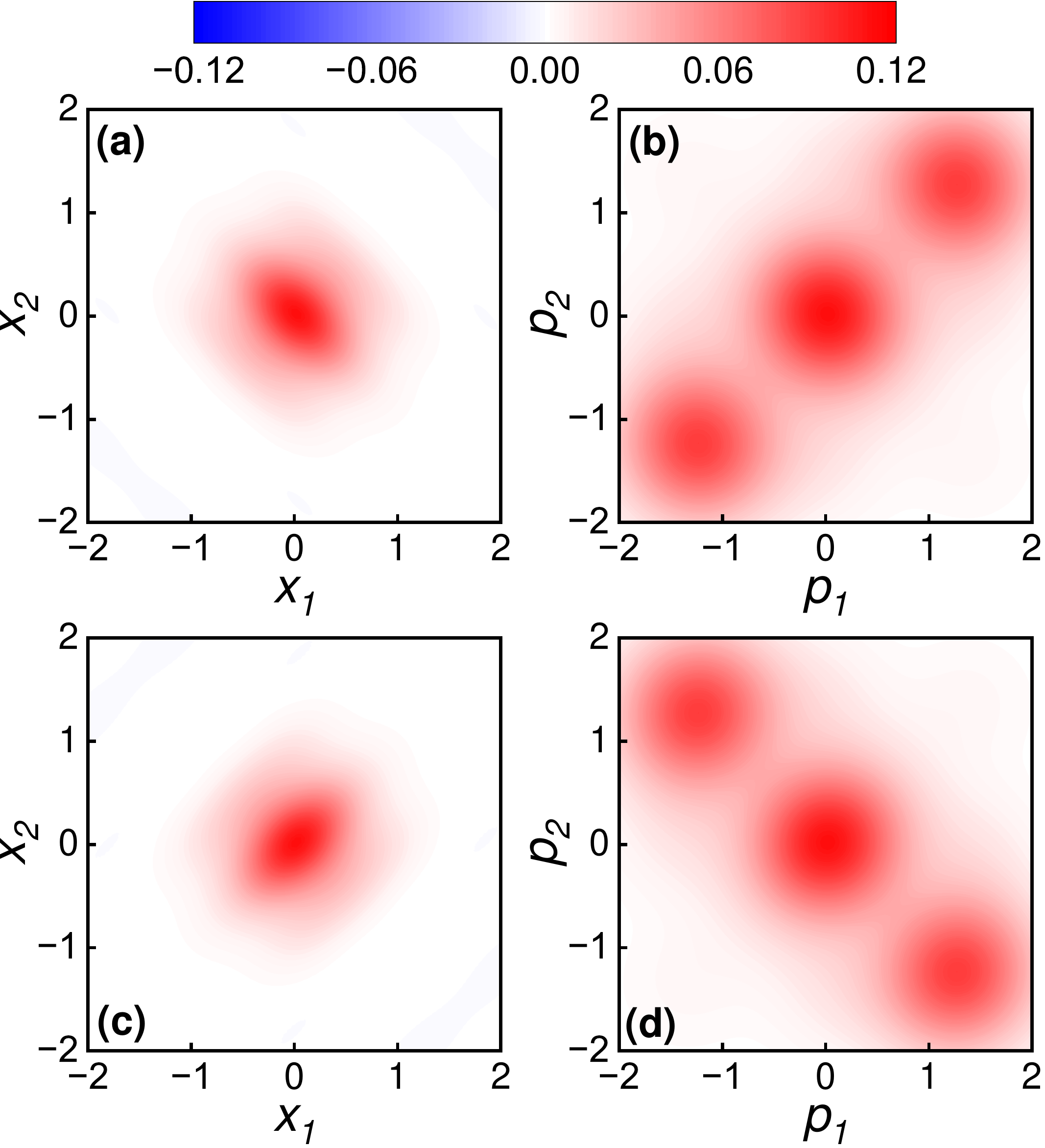}
\caption{The joint Wigner function for the two magnon modes. The numerical simulation parameters are $N=15$ (Fock-space truncation), driving amplitude $s_a=1$, coupling strength $\lambda_d=0.8$, and Kerr non-linearity $K=0.75$.\ The evolution time is set to $t=20$ (in units of $s_a$).\ The subfigures display two-dimensional projections of the four-dimensional Wigner function:(a) and (b) projection onto the displacement and momentum quadratures when the phonon squeezing parameter is $s_b = 2.4i$. (c) and (d) Projection onto the displacement and momentum quadratures when the phonon squeezing parameter is $s_b = -2.4i$.} 
\label{fig5}
\end{figure}
\subsection{Numerical verification of the switchable dissipative coupling}
In this section, we show that the nonlinear Lindblad operators in Eq.~(\ref{contralcollective}) can realize the switching between two distinct dissipative couplings. The system under consideration is composed of three distinct parts: the magnon part, the collective dissipation part, and the mechanical part. The magnon part comprises two magnon modes, whose nonlinear pump and Kerr nonlinearity are described by Eqs.~(\ref{two-magnon}) and (\ref{two-kerr}), respectively. The dissipative coupling is achieved via the dissipation channels presented in Eq.~(\ref{contralcollective}). The mechanical mode is described by the Hamiltonian~(\ref{phonon_squeeze}) and the local loss~(\ref{lossofb}), which together can generate switchable squeezing. As demonstrated in the previous section and other related works \cite{10.1103/c91r-8t3h,12-scattering-magnon-bistable}, without the dissipative coupling, these two magnon modes will evolve into a separable two-mode cat state. However, under the influence of the dissipative coupling, the system evolves into an entangled cat state. Moreover, the type of dissipative coupling dictates the relative phase between these two magnon modes, which can be verified using the joint Wigner function~\cite{10.1126/science.aaf2941}.  
\begin{figure}[t]
\center
\includegraphics[width=3in]{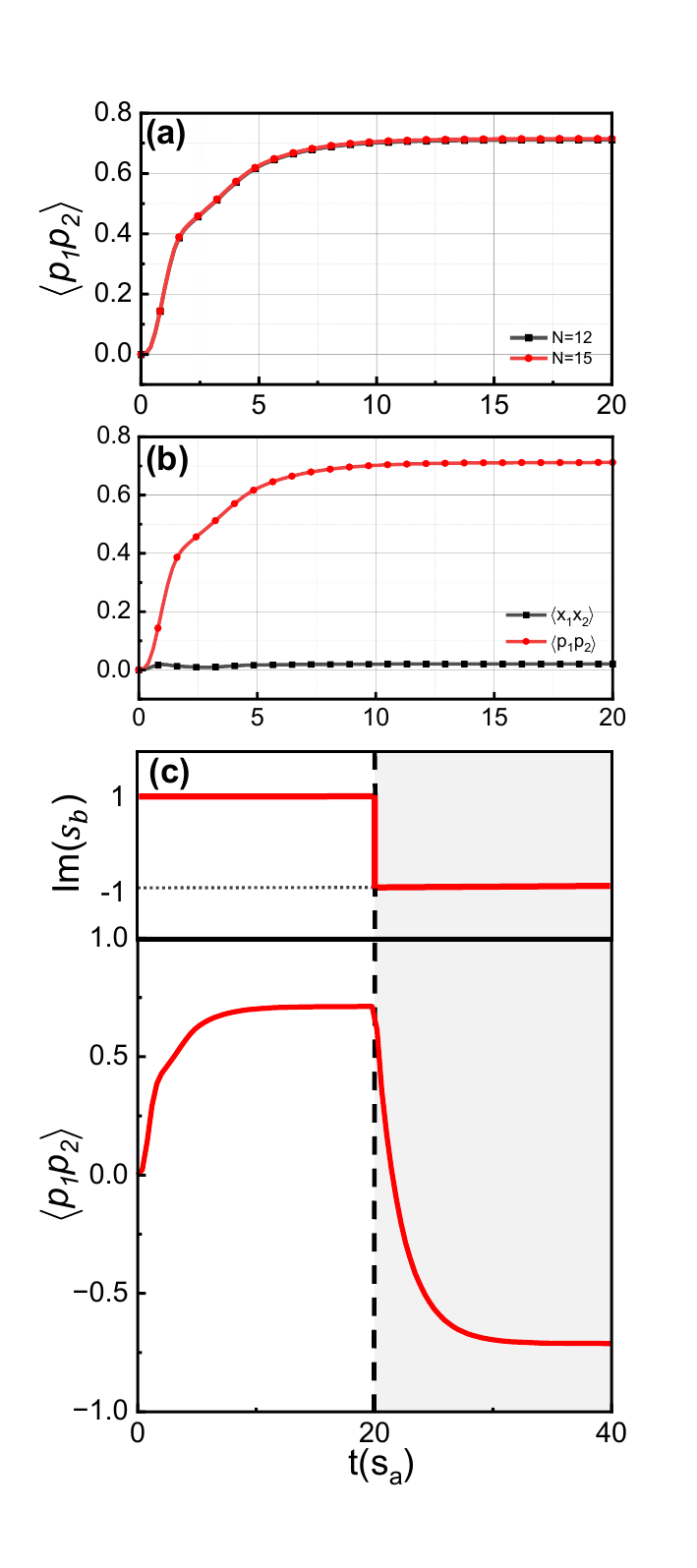}
\caption{Magnon mode quadrature evolution $\langle p_1 p_2 \rangle$ vs.time. Fixed parameters: $s_a=0.6$, $s_b=1i$, $\lambda_d=0.8$, and $K=0.75$. Time $t$ is in units of $s_a$.(a)$\langle p_1 p_2 \rangle$ comparison for $N=12$  and $N=15$.(b)Comparison of $\langle x_1 x_2 \rangle$ and $\langle p_1 p_2 \rangle$ for $N=15$.(c) $\operatorname{Im}(s_b)$ switches from $1i$ to $-1i$ at $t=20$ (Top), showing the $\langle p_1 p_2 \rangle$ phase transition response (Bottom).}
\label{fig6}
\end{figure}

We set the initial state to the vacuum state and numerically simulate the system's evolution using the Qutip package~\cite{10.1016/j.cpc.2012.02.021,10.1016/j.cpc.2012.11.019,2412.04705}. After a sufficiently long time evolution ($t=20s_a^{-1}$), we perform a partial trace over the mechanical mode and calculate the two-mode joint Wigner function~\cite{10.1126/science.aaf2941} of the two remaining magnon modes. It should be noted that the joint Wigner function is a function in four-dimensional space; consequently, only two important cross-sections of this Wigner function are presented in Fig.~\ref{fig5}. Owing to the phase effect of the nonlinear pump in the magnon modes, the position cross-section of the joint Wigner function corresponds to the interference pattern, whereas the momentum cross-section corresponds to the two steady states of the magnon mode. In Figs.~\ref{fig5}(a) and (b), we set the squeezing parameter of the mechanical mode to $s_b=2.4i$, so that the position quadrature is squeezed. Such position squeezing can suppress the loss channel $L_2$ in Eq.~(\ref{contralcollective}) while enhancing $L_1$.

Although the joint Wigner function provides an intuitive visualization of the system, it is not suitable for characterizing its dynamical evolution. As a result, we use the correlation function of the magnon modes to illustrate the effect of the switchable dissipative coupling. Since the two components of the magnon mode reside in the momentum quadrature, as illustrated in Fig.~\ref{fig3}, we use the momentum correlation function to characterize the relative phase between the two magnon modes. The time evolution of the correlation function of the dimensionless momentum $\langle p_1p_2\rangle$ defined in Eq.~(\ref{dimensionlessquadrature}) is shown in Fig.~\ref{fig6}. 

Figure~\ref{fig6}(a) displays the time evolution of the momentum correlation function for different Hilbert space truncation dimensions. Specifically, the black curve corresponds to truncation $N=12$, and the red curve corresponds to $N=15$. The results obtained for $N = 12$ are in complete agreement with those for $N = 15$, indicating that a truncation dimension of $N=12$ is sufficient. In addition, under the influence of a ferromagnetic dissipative coupling ${\rm Im}(s_b)>0$, the time evolution of the correlation function is clearly observed to transition from its initial value of $0$ to a steady-state value of approximately $0.7$. Fig.~\ref{fig6}(b) presents a comparison between the position correlation function (black curve) and the momentum correlation function (red curve). The position correlation function is negligible compared to the momentum correlation function, which validates our previous picture of effective spins being encoded in the momentum quadrature.

To illustrate the dynamical switching of the dissipative coupling, we consider a time-dependent nonlinear pump applied to the mechanical mode. This pump is initially set to ${\rm Im}(s_b)>0$ and is subsequently switched to ${\rm Im}(s_b)<0$, as depicted in the upper part of Fig.~\ref{fig6}(c). The lower part of Fig.~\ref{fig6}(c) displays the time evolution of the momentum correlation function. The evolution prior to the switching of $s_b$ is consistent with the results presented in Figs.~\ref{fig6}(a) and (b),i.e., the correlation function gradually approaches a steady positive value. Following the sign reversal of $s_b$, the correlation function begins to evolve toward a negative value, which corresponds to an antiferromagnetic alignment of the two optical spins.\ Therefore, our proposal enables the dynamical switching between different types of dissipative coupling.
\begin{figure}[t]
\center
\includegraphics[width=3in]{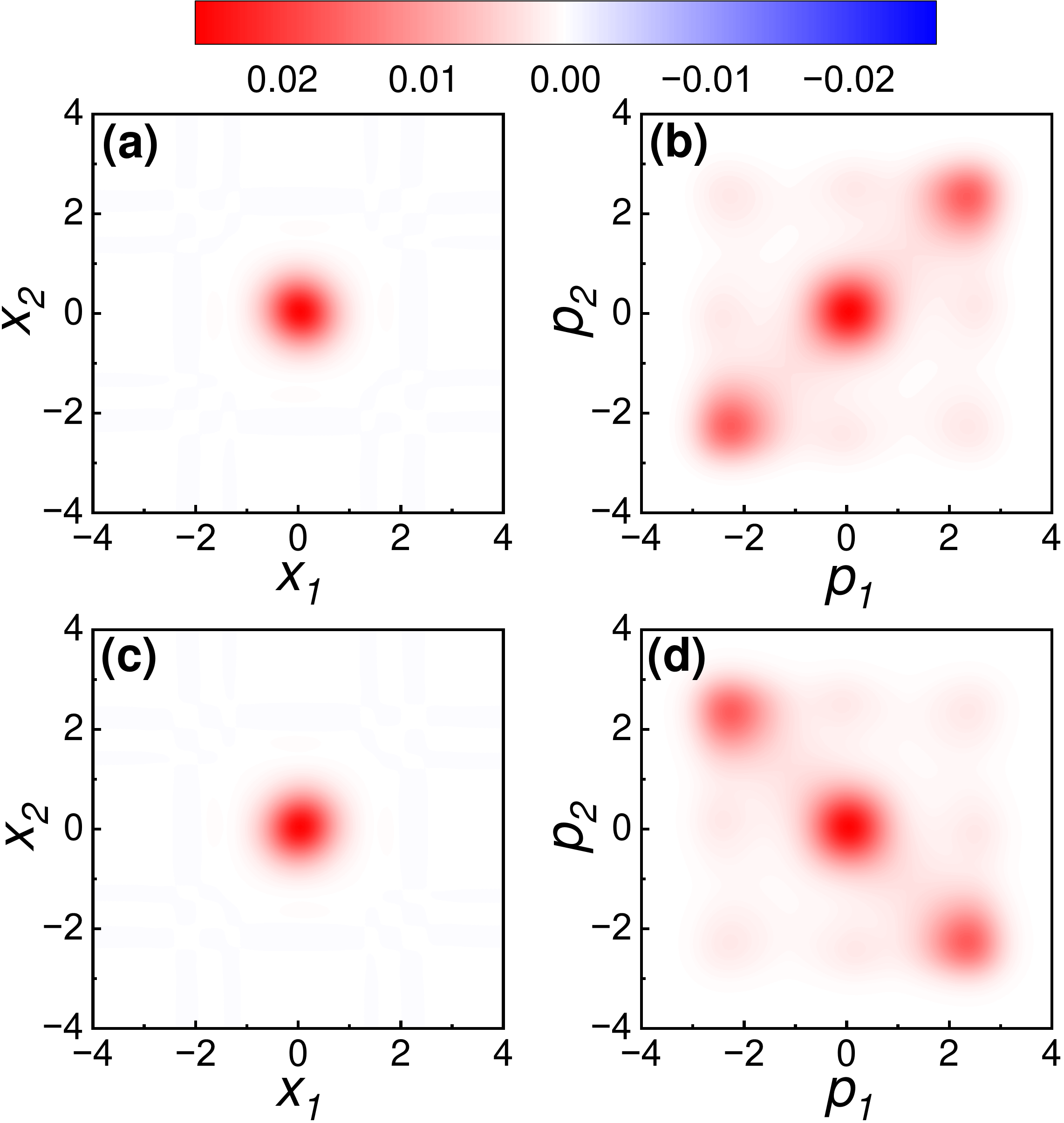}
\caption{The joint Wigner function projections of the magnon mode under loss and squeezing.
Fixed parameters are $N=12$, $K=0.75$, $\lambda_d=0.8$, $s_a=4.2$ and $\lambda_m=2$. Evolution time $t=20$ (in units of $s_a$).(a) Displacement projection and (b) Momentum projection for $s_b=2.4i$.(c) Displacement projection and (d) Momentum projection for $s_b=-2.4i$.}
\label{fig7}
\end{figure}
\subsection{Robustness of the switchable loss channels}
In the preceding section, we demonstrated the effects of switchable collective loss using the two-mode joint Wigner function and the momentum correlation function. We now consider the robustness of our proposal against other unwanted dissipation mechanisms. Note that the three-body coupling~\cite{santiouhe} is the first-order perturbation of the magnon-cavity coupling with respect to the mechanical mode. As a result, the zeroth order terms, which remain unmodified by the mechanical mode, can also contribute to the magnon loss.\ To investigate the influence of these zeroth-order contributions, we incorporate the following two loss channels,
\begin{equation}
\begin{aligned}
L_3 &= -\lambda_m({a}_1+{a}_2)\\
L_4 &= -\lambda_m({a}_1-{a}_2),
\end{aligned}
\end{equation}
with a significantly stronger loss rate $\Gamma_{\rm ncc}=4s_a$ compared to the switchable part $\Gamma_{\rm cc}=0.64s_a$. To compensate the additional loss, we introduce enhanced pumps into the magnon modes,
\begin{eqnarray}\label{two-magnon-withloss}
\begin{aligned}
    \tilde{H}_{\text{s}} =&\tilde{s}_a^*{a_1}^\dagger{a}_1^\dagger+\tilde{s}_a{a}_1{a}_1+\tilde{s}_a^*{a}_2^\dagger{a}_2^\dagger+\tilde{s}_a{a}_2{a}_2,
\end{aligned}
\end{eqnarray}
with $\tilde{s}_a=4.2s_a$. All other parameters remain consistent with those used in Figs.~\ref{fig5} and \ref{fig6}. 

We first provide an intuitive picture by utilizing the joint Wigner function of the reduced magnon state, as illustrated in Fig.~\ref{fig7}. Figure \ref{fig7} shows typical joint Wigner functions of catlike two-mode states, which can represent two correlated effective spins. In Figs.~\ref{fig7}(a) and (b), the position quadrature of the mechanical mode is squeezed, consequently, the coupling assumes a ferromagnetic form. The two nonzero components in Fig.~\ref{fig7}(b) exhibit positive momentum correlations, which agrees with the expected effects of a ferromagnetic dissipative coupling. In Figs.~\ref{fig7}(c) and (d), the coupling is set to the antiferromagnetic form, and the two nonzero components in Fig.~\ref{fig7}(d) exhibit negative momentum correlations. These results prove that the switchable dissipative coupling is still effective even under the influence of strong dissipation mechanisms that cannot be modified by the mechanical mode.

\begin{figure}[t]
\center
\includegraphics[width=3in]{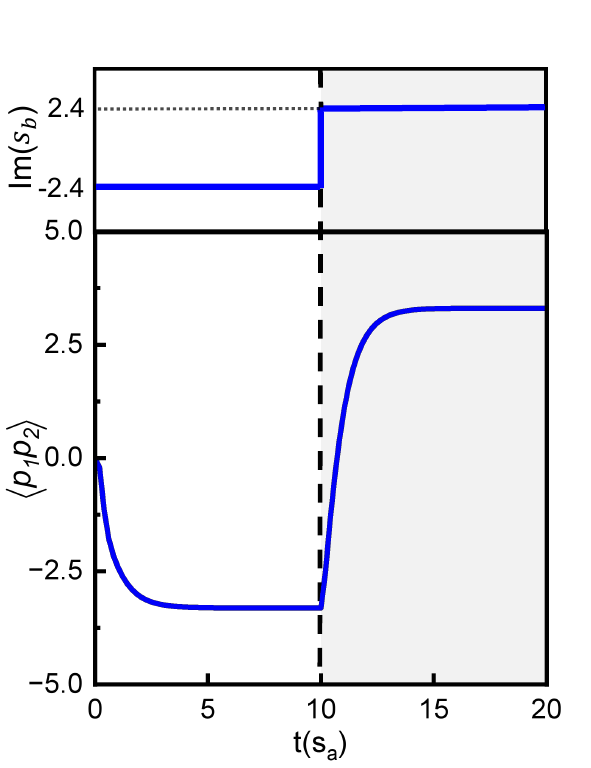}
\caption{Evolution of the magnon momentum quadrature expectation value with changable parameter $\operatorname{Im}(s_b)$ from $-2.4i$ to $2.4i$ when $N=12,\lambda_d=0.8, K = 0.75, s_a=4.2,\lambda_m=2$}
\label{fig8} 
\end{figure}

Following the approach of the preceding section, we employ the momentum correlation function to demonstrate the dynamical switching of the coupling. In Fig.~\ref{fig8}, the imaginary part of $s_b$ is first set to a negative value and is subsequently switched to a positive value after a period of time ($10s_a^{-1}$). The evolution of the momentum correlation function is consistent with the switching of the coupling. Therefore, the dynamical switching can also be achieved even under the influence of strong constant dissipation.

\section{Conclusions}
\label{conclusion}
We proposed a scheme to realize switchable dissipative coupling in magnon systems, based on three-body coupling, which is crucial for applying these systems in solving combinatorial optimization problems. The switchable dissipative coupling consists of two distinct collective dissipation channels induced by two open conducting wires, where the coupling to the magnon modes is modulated by a mechanical mode (i.e., three-body coupling). By modulating the phase of the nonlinear pump in the mechanical mode, the squeezed quadrature can be tuned to control the active dissipation channel. We numerically verified the effects of our proposal using two magnon modes coupled by the switchable dissipative coupling. The numerical results demonstrate that different control parameters in the mechanical mode lead to distinct relative phases between the two magnon modes, consistent with the design of the switchable coupling. Furthermore, switching during the dynamical evolution is also feasible. In addition, we investigated the influence of unwanted dissipation arising from the constant term of the magnon-wire coupling, which can be effectively compensated by incorporating stronger dissipation in the magnon modes. 

Our work provides a potential pathway for the application of magnon systems in Ising machines for solving complex combinatorial optimization problems. These switchable couplings may also improve the performance of magnon systems in quantum computing and quantum communications. 
\section{Acknowledgments}
This research is supported by the National Natural Science Foundation of China (NSFC) (Grant No.~12405028); Zhejiang Provincial Natural Science Foundation of China under Grant No.~LQ24A050002; the Innovation Program for Quantum Science and Technology (2023ZD0300904). A.X.C. is supported by the National Natural Science Foundation of China (Grants No.12575031). 

\appendix
\section{single-mode loss of the three-body system}
In this research, we want to use loss channels to manipulate the quantum state of magnon mode. So we achieve differential dissipation strengths through the coupling within the tripartite system. The interaction Hamiltonian of the three-body system is given by:
\begin{eqnarray}
    \begin{aligned}   
    H_{\text{int}} &= \lambda({b}+{b}^\dagger)({a}{c}{^\dagger}+{a}^\dagger{c})\\&=\lambda({b}+{b}^\dagger){a}{c}{^\dagger}+\lambda({b}+{b}^\dagger){a}^\dagger{c}.
\end{aligned}
\end{eqnarray}
To address the squeezing control terms, we first consider eliminating this interactive Hamiltonian. In this system, ${c}$ is the only operator with dissipation. Consequently, the Langevin equation for each operator is expressed as: 
\begin{eqnarray}
\begin{aligned}
     \dot{{a}} &= i[H_{\text{int}},{a}] = -i\lambda({b}+{b}{^\dagger}){c},\\\dot{{b}} &= i[H_{\text{int}},{b}] = -i\lambda({a}{^\dagger}{c}+{a}{c}^\dagger),\\\dot{{c}} &= i[H_{\text{int}},{c}] = -i\lambda({b}+{b}{^\dagger}){a} - \gamma_{c}{c}+\sqrt{2\gamma{_c}}C{_{in}}.
\end{aligned}
\end{eqnarray}
We suppose the total system reach equilibrium($\dot{{c}}=0$) and zero-mean noise input($C_{in}=0$). 
\begin{equation}
\begin{aligned}
     \dot{{c}} &= i[H_{\text{int}},{c}] \\&= -i\lambda({b}+{b}{^\dagger}){a} - \gamma_{c}{c}+\sqrt{2\gamma{_c}}C{_{in}}=0,
\end{aligned}
\end{equation}
so we get: 
\begin{equation}
    {c} = \frac{-i\lambda({b}+{b}{^\dagger}){a}+C_{in}}{\gamma_{c}}.   
\end{equation}
Meanwhile,
\begin{equation}
\begin{aligned}
     \dot{{a}} &= -i\lambda({b}+{b}{^\dagger}){c}\\&=-i\lambda({b}+{b}{^\dagger})\frac{-i\lambda({b}+{b}{^\dagger}){a}+C_{in}}{\gamma_{c}}  \\&=\frac{-\lambda{^2}({b}+{b}{^\dagger})^{2}{a}}{{\gamma_{c}}}+\frac{{-\lambda{^2}({b}+{b}{^\dagger})C_{in}}}{\gamma_{c}},
\end{aligned}
\end{equation}
we obtain the equation of motion for $\sigma{_m}{^-}$ and subsequently harness the Master Equation formalism to describe the Liouvillian operator $\hat{L}$.
\begin{equation}
\begin{aligned}
    {\dot\rho} &= i[H,\rho] -\frac{\Gamma}{2}(2L\rho L^\dagger - L^\dagger L\rho - \rho L^\dagger L),\\<\dot{{a}}> =&<\dot{\rho}{a}> \\=&i<[H,\rho]{a}> - \frac{r}{2}[<2L\rho L^\dagger{a}> \\&- <L^\dagger L\rho{a}> - <\rho L^\dagger L{a}>]\\=&i<[H,\rho]{a}>\\&- \frac{\Gamma}{2}[<\rho L[L^\dagger,{a}]>+<\rho L^\dagger[{a},L]>].
\end{aligned}
\end{equation}
Then through $\Gamma L[L^\dagger,{a}] == -\frac{\lambda{^2}}{\gamma_c}({b}+{b}{^\dagger})^{2}{a}$, we can calculate:
\begin{equation}
    L = -\lambda_s({b}+{b}{^\dagger}){a},
\end{equation}
with the induced loss rate $\Gamma = \frac{\lambda^2}{\gamma_c}$, which can be manipulated.

\section{collective dissipation of the triple-body system}
In this appendix, we briefly discuss the collective loss channels within the tripartite coupling system. The interaction Hamiltonian for the system is given by:
\begin{equation}
\begin{aligned}
    H_{\text{int,1}}  = &\lambda_c({b}+{b}^\dagger)({a}_1{c}{^\dagger}+{a}{_1}{^\dagger}{c})\\&+\lambda_c({b}+{b}^\dagger)({a}_2{c}{^\dagger}+{a}{_2}{^\dagger}{c})\\= &\lambda_c({b}+{b}^\dagger)[({a}_1+{a}_2){c}{^\dagger}+({a}_1+{a}_2)^\dagger {c}].
\end{aligned}
\end{equation}
Analogous to the single-mode treatment, the Langevin equations for each operator can be described as,
\begin{equation}
\begin{aligned}
     \dot{{a}}_1 &= i[H_{\text{int,1}},{a}_1] = -i\lambda_c({b}+{b}{^\dagger}){c},\\\dot{{a}}_2 &= i[H_{\text{int,1}},{a}_2] = -i\lambda_c({b}+{b}{^\dagger}){c},\\
     \dot{{b}} &= i[H_{\text{int,1}},{b}] = -i\lambda_c[({a}_1+{a}_2){c}+({a}_1+{a}_2)^\dagger {c}),\\\dot{{c}} &= i[H_{\text{int,1}},{c}] \\&= -i\lambda_c({b}+{b}{^\dagger})({a}_1+{a}_2)  - \gamma_{c}{c}+\sqrt{2\gamma{_c}}C{_{in}}.
\end{aligned}
\end{equation}
We assume that the coupled system is in thermal equilibrium.
\begin{equation}
\begin{aligned}
     \dot{{c}} &= i[H_{\text{int,1}},{c}] \\&= -i\lambda_c({b}+{b}{^\dagger})({a}_1+{a}_2) - \gamma_{c}{c}+\sqrt{2\gamma{_c}}C{_{in}}\\&=0,
\end{aligned}
\end{equation}
so we get: 
\begin{equation}
    {c} = \frac{-i\lambda_c({b}+{b}{^\dagger})({a}_1+{a}_2)}{\gamma_{c}}.  
\end{equation}
Take this result back into the Langevin equation of ${a}_1+{a}_2$:
\begin{equation}
\begin{aligned}
\dot{a}_1+\dot{a}_2&=-2i\lambda_c({b}+{b}{^\dagger}){c}\\&=-2\frac{\lambda_c{^2}({b}+{b}{^\dagger})^2({a}_1+{a}_2)}{\gamma_{c}}
\end{aligned}
\end{equation}
By employing the Master Equation formalism and calculating the expectation value of ${a}_1+{a}_2$, we derive the effective Lindblad operator corresponding to $H_{int1}$.
\begin{equation}
\begin{aligned}
   <\dot{a}_1+\dot{a}_2> = &<\dot{\rho}({a}_1+{a}_2)> \\=&i<[H,\rho]({a}_1+{a}_2)> \\&-\frac{\Gamma_{\rm cc}}{2}[<2L\rho L^\dagger ({a}_1+{a}_2)> \\&- <L^\dagger L\rho({a}_1+{a}_2)> - <\rho L^\dagger L({a}_1+{a}_2)>]\\=&i<[H,\rho]({a}_1+{a}_2)> \\&-\frac{\Gamma_{\rm cc}}{2}[<\rho L[L^\dagger,({a}_1+{a}_2)]>\\&+<\rho L^\dagger[({a}_1+{a}_2),L]>].
\end{aligned}
\end{equation}
Through $\Gamma_{\rm cc}L[L^\dagger,{a}{_1}+{a}_2] = -\frac{\lambda_c{^2}}{\gamma_c}({b}+{b}{^\dagger})^{2}({a}{_1}+{a}_2)$, we can calculate:
\begin{equation}
    L_1 = -({b}+{b}{^\dagger})({a}{_1}+{a}_2),
\end{equation}
with the controllable collective loss rate $\Gamma_{\rm cc} = \frac{\lambda_c^2}{\gamma_c}$. 

Note that the type of coupling can be adjusted by controlling the relative position of the YIG sphere and the conducting line. The bosonic modes in a 1-D conducting line or in a semi-1-D cantilever can be expressed as:
\begin{eqnarray}
A(r,t)=C(e^{ikr-i\omega t}a+{\rm H.c.}).
\end{eqnarray}
Here, $a$ is the annihilation operator of the bosonic mode; $k$ is the wave vector; $\omega$ is the frequency; $r$ is the position and $C$ is the normalization factor for the Fourier transformation. By introducing another conducting line at a different position, it is possible to introduce a different three-body coupling term,  
\begin{equation}
\begin{aligned}
    H_{\text{int2}} & = i\lambda_c({b}-{b}^\dagger)({a}_1{c}{^\dagger}-{a}{_1}{^\dagger}{c})*j+\lambda_c({b}-{b}^\dagger)({a}{_2}{^\dagger}{c}-{a}_2{c}{^\dagger})\\&= i\lambda_c({b}-{b}^\dagger)[({a}_1-{a}_2){c}{^\dagger}+({a}_2-{a}_1)^\dagger{c}],
\end{aligned}
\end{equation}
For this part, the Linblod operator after elimination can be described as,                 
\begin{equation}
    L_2 = -\lambda_d({b}-{b}{^+})({a}{_1}-{a}_2)i,
\end{equation}
so we can get our aim state by controlling the two Linblad operators($ L_1,L_2$).

\vspace{12pt}
\bibliography{pref}

@article{1magnonmaterial1,
  title = {Magnon Dirac materials},
  author = {Fransson, J. and Black-Schaffer, A. M. and Balatsky, A. V.},
  journal = {Phys. Rev. B},
  volume = {94},
  issue = {7},
  pages = {075401},
  numpages = {6},
  year = {2016},
  month = {Aug},
  publisher = {American Physical Society},
  doi = {10.1103/PhysRevB.94.075401},
  url = {https://link.aps.org/doi/10.1103/PhysRevB.94.075401}
}

@article{1magnonmaterial2, 
    title={Magnon transistor for all-magnon data processing}, 
    volume={5}, 
    ISSN={2041-1723}, 
    DOI={10.1038/ncomms5700}, 
    number={1}, 
    journal={Nature Communications}, 
    publisher={Nature Communications}, 
    author={Chumak, Andrii V. and Serga, Alexander A. and Hillebrands, Burkard}, 
    year={2014}, 
    pages={4700},
    url={https://dx.doi.org/10.1038/ncomms5700}, 
}

@article{1magnonmaterial3, 
    title={Introduction to antiferromagnetic magnons}, 
    volume={126}, 
    ISSN={0021-8979}, 
    DOI={10.1063/1.5109132}, 
    number={15}, 
    journal={Journal of Applied Physics}, 
    publisher={Journal of Applied Physics}, 
    author={Rezende, Sergio M. and Azevedo, Antonio and Rodríguez-Suárez, Roberto L.}, 
    year={2019}, 
    pages={151101} 
}

@article{1magnonmaterial4, 
    title={Colloquium: Spontaneous magnon decays}, 
    volume={85}, 
    ISSN={0034-6861}, 
    DOI={10.1103/revmodphys.85.219}, 
    number={1}, 
    journal={Reviews of Modern Physics}, 
    publisher={Reviews of Modern Physics}, 
    author={Zhitomirsky, M. E. and Chernyshev, A. L.}, 
    year={2013}, 
    pages={219–242} 
}

@article{2magnonnonlinearity1, 
    title={Emergent coherent modes in nonlinear magnonic waveguides detected at ultrahigh frequency resolution}, 
    author={An, K. and Xu, M. and Mucchietto, A. and Kim, C. and Moon, K.-W. and Hwang, C. and Grundler, D.}, 
    volume={15}, 
    DOI={10.1038/s41467-024-51483-7}, 
    number={1}, 
    journal={Nature Communications}, 
    publisher={Nature Communications}, 
    ISSN={2041-1723}, 
    pages = {7302},
    year={2024},
    url = {https://doi.org/10.1038/s41467-024-51483-7}
}

@article{2magnonnonlinearity2,
    title={Nonlinear losses in magnon transport due to four-magnon scattering}, 
    volume={117}, 
    ISSN={0003-6951}, 
    DOI={10.1063/5.0015269}, 
    number={4}, 
    journal={Applied Physics Letters}, 
    publisher={Applied Physics Letters}, 
    author={Hula, Tobias and Schultheiss, Katrin and Buzdakov, Aleksandr and Körber, Lukas and Bejarano, Mauricio and Flacke, Luis and Liensberger, Lukas and Weiler, Mathias and Shaw, Justin M. and Nembach, Hans T. and et al.}, 
    year={2020}, 
    pages={042404} 
}

@article{2magnonnonlinearity3,
  title = {Detection sensitivity enhancement of magnon Kerr nonlinearity in cavity magnonics induced by coherent perfect absorption},
  author = {Zhang, Guo-Qiang and Wang, Yimin and Xiong, Wei},
  journal = {Phys. Rev. B},
  volume = {107},
  issue = {6},
  pages = {064417},
  numpages = {9},
  year = {2023},
  month = {Feb},
  publisher = {American Physical Society},
  doi = {10.1103/PhysRevB.107.064417},
  url = {https://link.aps.org/doi/10.1103/PhysRevB.107.064417}
}

@article{3magnonscattering1,
  title = {Direct observation of nonlinear four-magnon scattering in spin-wave microconduits},
  author = {Schultheiss, H. and Vogt, K. and Hillebrands, B.},
  journal = {Phys. Rev. B},
  volume = {86},
  issue = {5},
  pages = {054414},
  numpages = {4},
  year = {2012},
  month = {Aug},
  publisher = {American Physical Society},
  doi = {10.1103/PhysRevB.86.054414},
  url = {https://link.aps.org/doi/10.1103/PhysRevB.86.054414}
}

@article{3magnonscattering2,
  title = {Two magnon scattering in ultrathin ferromagnets: The case where the magnetization is out of plane},
  author = {Landeros, P. and Arias, Rodrigo E. and Mills, D. L.},
  journal = {Phys. Rev. B},
  volume = {77},
  issue = {21},
  pages = {214405},
  numpages = {13},
  year = {2008},
  month = {Jun},
  publisher = {American Physical Society},
  doi = {10.1103/PhysRevB.77.214405},
  url = {https://link.aps.org/doi/10.1103/PhysRevB.77.214405}
}

@article{3magnonscattering3,
    author = {Heinrich, B. and Cochran, J. F. and Hasegawa, R.},
    title = {FMR linebroadening in metals due to two‐magnon scattering},
    journal = {Journal of Applied Physics},
    volume = {57},
    number = {8},
    pages = {3690-3692},
    year = {1985},
    month = {04},
    issn = {0021-8979},
    doi = {10.1063/1.334991},
    url = {https://doi.org/10.1063/1.334991},
}

@article{4magnonspintronics1,
  title={Magnon spintronics},
  author={Chumak, Andrii V and Vasyuchka, Vitaliy I and Serga, Alexander A and Hillebrands, Burkard},
  journal={Nature physics},
  volume={11},
  number={6},
  pages={453--461},
  year={2015},
  publisher={Nature Publishing Group UK London},
  doi = {10.1038/nphys3341},
}

@article{4magnonspintronics2,
    title = {Quantum magnonics: When magnon spintronics meets quantum information science},
    journal = {Physics Reports},
    volume = {965},
    pages = {1-74},
    year = {2022},
    issn = {0370-1573},
    doi = {https://doi.org/10.1016/j.physrep.2022.03.002},
    url = {https://www.sciencedirect.com/science/article/pii/S0370157322000977},
    author = {H.Y. Yuan and Yunshan Cao and Akashdeep Kamra and Rembert A. Duine and Peng Yan},
}

@article{4magnonspintronics3,
    title = {Parallel pumping for magnon spintronics: Amplification and manipulation of magnon spin currents on the micron-scale},
    journal = {Physics Reports},
    volume = {699},
    pages = {1-34},
    year = {2017},
    issn = {0370-1573},
    doi = {https://doi.org/10.1016/j.physrep.2017.07.003},
    url = {https://www.sciencedirect.com/science/article/pii/S0370157317302004},
    author = {T. Brächer and P. Pirro and B. Hillebrands},
}

@article{5hybrid-magnon-system1,
    title = {Hybridizing Ferromagnetic Magnons and Microwave Photons in the Quantum Limit},
    author = {Tabuchi, Yutaka and Ishino, Seiichiro and Ishikawa, Toyofumi and Yamazaki, Rekishu and Usami, Koji and Nakamura, Yasunobu},
    journal = {Phys. Rev. Lett.},
    volume = {113},
    issue = {8},
    pages = {083603},
    numpages = {5},
    year = {2014},
    month = {Aug},
    publisher = {American Physical Society},
    doi = {10.1103/PhysRevLett.113.083603},
    url = {https://link.aps.org/doi/10.1103/PhysRevLett.113.083603}
}

@article{5hybrid-magnon-system2,
    title = {Electron-Magnon Interaction in Ferromagnetic Semiconductors},
    author = {Woolsey, Roy B. and White, Robert M.},
    journal = {Phys. Rev. B},
    volume = {1},
    issue = {11},
    pages = {4474--4486},
    numpages = {0},
    year = {1970},
    month = {Jun},
    publisher = {American Physical Society},
    doi = {10.1103/PhysRevB.1.4474},
    url = {https://link.aps.org/doi/10.1103/PhysRevB.1.4474}
}

@article{5hybrid-magnon-system3,
  title = {Nonreciprocal entanglement in cavity-magnon optomechanics},
  author = {Chen, Jiaojiao and Fan, Xiao-Gang and Xiong, Wei and Wang, Dong and Ye, Liu},
  journal = {Phys. Rev. B},
  volume = {108},
  issue = {2},
  pages = {024105},
  numpages = {8},
  year = {2023},
  month = {Jul},
  publisher = {American Physical Society},
  doi = {10.1103/PhysRevB.108.024105},
  url = {https://link.aps.org/doi/10.1103/PhysRevB.108.024105}
}

@article{5hybrid-magnon-system4,
    doi = {10.7567/1882-0786/ab248d},
    url = {https://dx.doi.org/10.7567/1882-0786/ab248d},
    year = {2019},
    month = {jun},
    publisher = {IOP Publishing},
    volume = {12},
    number = {7},
    pages = {070101},
    author = {Lachance-Quirion, Dany and Tabuchi, Yutaka and Gloppe, Arnaud and Usami, Koji and Nakamura, Yasunobu},
    title = {Hybrid quantum systems based on magnonics},
    journal = {Applied Physics Express},
}

@article{6magnon-kerr-cat1,
  title = {Cat-state-like non-Gaussian entanglement in magnon systems},
  author = {Zhang, Zeyu and Gneiting, Clemens and Zhou, Zheng-Yang and Chen, Ai-Xi},
  journal = {Phys. Rev. A},
  volume = {111},
  issue = {3},
  pages = {033717},
  numpages = {11},
  year = {2025},
  month = {Mar},
  publisher = {American Physical Society},
  doi = {10.1103/PhysRevA.111.033717},
  url = {https://link.aps.org/doi/10.1103/PhysRevA.111.033717}
}

@article{6magnon-kerr-cat2,
  title = {Magnon cat states induced by photon parametric coupling},
  author = {Liu, Da-Wei and Wu, Ying and Si, Liu-Gang},
  journal = {Phys. Rev. Appl.},
  volume = {21},
  issue = {4},
  pages = {044018},
  numpages = {11},
  year = {2024},
  month = {Apr},
  publisher = {American Physical Society},
  doi = {10.1103/PhysRevApplied.21.044018},
  url = {https://link.aps.org/doi/10.1103/PhysRevApplied.21.044018}
}

@article{6magnon-kerr-cat3,
  title = {Remote Generation of Magnon Schr\"odinger Cat State via Magnon-Photon Entanglement},
  author = {Sun, Feng-Xiao and Zheng, Sha-Sha and Xiao, Yang and Gong, Qihuang and He, Qiongyi and Xia, Ke},
  journal = {Phys. Rev. Lett.},
  volume = {127},
  issue = {8},
  pages = {087203},
  numpages = {6},
  year = {2021},
  month = {Aug},
  publisher = {American Physical Society},
  doi = {10.1103/PhysRevLett.127.087203},
  url = {https://link.aps.org/doi/10.1103/PhysRevLett.127.087203}
}

@article{6magnon-kerr-cat4,
  title = {Magnon-Induced Nonreciprocity Based on the Magnon Kerr Effect},
  author = {Kong, Cui and Xiong, Hao and Wu, Ying},
  journal = {Phys. Rev. Appl.},
  volume = {12},
  issue = {3},
  pages = {034001},
  numpages = {9},
  year = {2019},
  month = {Sep},
  publisher = {American Physical Society},
  doi = {10.1103/PhysRevApplied.12.034001},
  url = {https://link.aps.org/doi/10.1103/PhysRevApplied.12.034001}
}

@article{7magnon-kerr-squeeze1,
    author = {Haghshenasfard, Zahra and Cottam, Michael G.},
    title = {Sub-Poissonian statistics and squeezing of magnons due to the Kerr effect in a hybrid coupled cavity–magnon system},
    journal = {Journal of Applied Physics},
    volume = {128},
    number = {3},
    pages = {033901},
    year = {2020},
    month = {07},
    issn = {0021-8979},
    doi = {10.1063/5.0012072},
    url = {https://doi.org/10.1063/5.0012072}
}

@article{7magnon-kerr-squeeze2, 
    title={Magnon-squeezing as a niche of quantum magnonics}, 
    volume={117}, 
    ISSN={0003-6951}, 
    DOI={10.1063/5.0021099}, 
    number={9}, 
    journal={Applied Physics Letters}, 
    publisher={Applied Physics Letters}, 
    author={Kamra, Akashdeep and Belzig, Wolfgang and Brataas, Arne}, 
    year={2020}, 
    pages={090501} 
}

@article{7magnon-kerr-squeeze3,
    title = {Kerr nonlinearity induced strong spin-magnon coupling},
    author = {Ji, Feng-Zhou and An, Jun-Hong},
    journal = {Phys. Rev. B},
    volume = {108},
    issue = {18},
    pages = {L180409},
    numpages = {7},
    year = {2023},
    month = {Nov},
    publisher = {American Physical Society},
    doi = {10.1103/PhysRevB.108.L180409},
    url = {https://link.aps.org/doi/10.1103/PhysRevB.108.L180409}
}

@article{8kerr-decoherence2,
title = {Coherent manipulation of the surface plasmon resonance sensing at the dielectric-graphene interface under Cross-Kerr nonlinearity effect},
journal = {Journal of Magnetism and Magnetic Materials},
volume = {618},
pages = {172858},
year = {2025},
issn = {0304-8853},
doi = {https://doi.org/10.1016/j.jmmm.2025.172858},
url = {https://www.sciencedirect.com/science/article/pii/S0304885325000897},
author = {Qaisar Khan and Mostafa R. Abukhadra and Ahmed M. El-Sherbeeny and Jeong Ryeol Choi and Asghar Ali and Majid Khan},
}

@article{9YIG-kerr2,
  title = {Magneto-Optical Kerr Effects of Yttrium-Iron Garnet Thin Films Incorporating Gold Nanoparticles},
  author = {Tomita, Satoshi and Kato, Takeshi and Tsunashima, Shigeru and Iwata, Satoshi and Fujii, Minoru and Hayashi, Shinji},
  journal = {Phys. Rev. Lett.},
  volume = {96},
  issue = {16},
  pages = {167402},
  numpages = {4},
  year = {2006},
  month = {Apr},
  publisher = {American Physical Society},
  doi = {10.1103/PhysRevLett.96.167402},
  url = {https://link.aps.org/doi/10.1103/PhysRevLett.96.167402}
}

@article{9YIG-kerr3,
  title = {Magnon Kerr effect in a strongly coupled cavity-magnon system},
  author = {Wang, Yi-Pu and Zhang, Guo-Qiang and Zhang, Dengke and Luo, Xiao-Qing and Xiong, Wei and Wang, Shuai-Peng and Li, Tie-Fu and Hu, C.-M. and You, J. Q.},
  journal = {Phys. Rev. B},
  volume = {94},
  issue = {22},
  pages = {224410},
  numpages = {8},
  year = {2016},
  month = {Dec},
  publisher = {American Physical Society},
  doi = {10.1103/PhysRevB.94.224410},
  url = {https://link.aps.org/doi/10.1103/PhysRevB.94.224410}
}

@article{10YIG-in-QED1,
  title = {Nonlinear pumping induced multipartite entanglement in a hybrid magnon cavity QED system},
  author = {Zhou, Y. and Xie, S. Y. and Zhu, C. J. and Yang, Y. P.},
  journal = {Phys. Rev. B},
  volume = {106},
  issue = {22},
  pages = {224404},
  numpages = {7},
  year = {2022},
  month = {Dec},
  publisher = {American Physical Society},
  doi = {10.1103/PhysRevB.106.224404},
  url = {https://link.aps.org/doi/10.1103/PhysRevB.106.224404}
}

@article{10YIG-in-QED2,
  title = {High-Cooperativity Cavity QED with Magnons at Microwave Frequencies},
  author = {Goryachev, Maxim and Farr, Warrick G. and Creedon, Daniel L. and Fan, Yaohui and Kostylev, Mikhail and Tobar, Michael E.},
  journal = {Phys. Rev. Appl.},
  volume = {2},
  issue = {5},
  pages = {054002},
  numpages = {11},
  year = {2014},
  month = {Nov},
  publisher = {American Physical Society},
  doi = {10.1103/PhysRevApplied.2.054002},
  url = {https://link.aps.org/doi/10.1103/PhysRevApplied.2.054002}
}

@article{11magnon-bistable1,
    author = {Makiuchi, Takahiko and Hioki, Tomosato and Shimazu, Yoshiki and Oikawa, Yasuyuki and Yokoi, Naoto and Daimon, Shunsuke and Saitoh, Eiji},
    title = {Parametron on magnetic dot: Stable and stochastic operation},
    journal = {Applied Physics Letters},
    volume = {118},
    number = {2},
    pages = {022402},
    year = {2021},
    month = {01},
    issn = {0003-6951},
    doi = {10.1063/5.0038946},
    url = {https://doi.org/10.1063/5.0038946}
}

@article{11magnon-bistable3,
  title = {Bistability of Cavity Magnon Polaritons},
  author = {Wang, Yi-Pu and Zhang, Guo-Qiang and Zhang, Dengke and Li, Tie-Fu and Hu, C.-M. and You, J. Q.},
  journal = {Phys. Rev. Lett.},
  volume = {120},
  issue = {5},
  pages = {057202},
  numpages = {6},
  year = {2018},
  month = {Jan},
  publisher = {American Physical Society},
  doi = {10.1103/PhysRevLett.120.057202},
  url = {https://link.aps.org/doi/10.1103/PhysRevLett.120.057202}
}

@article{11magnon-bistable4,
  title = {Bistability in dissipatively coupled cavity magnonics},
  author = {Pan, H. and Yang, Y. and An, Z. H. and Hu, C.-M.},
  journal = {Phys. Rev. B},
  volume = {106},
  issue = {5},
  pages = {054425},
  numpages = {11},
  year = {2022},
  month = {Aug},
  publisher = {American Physical Society},
  doi = {10.1103/PhysRevB.106.054425},
  url = {https://link.aps.org/doi/10.1103/PhysRevB.106.054425}
}

@article{11magnon-bistable5,
  title = {Bistability of squeezing and entanglement in cavity magnonics},
  author = {Yang, Zhi-Bo and Jin, Hua and Jin, Jing-Wen and Liu, Jian-Yu and Liu, Hong-Yu and Yang, Rong-Can},
  journal = {Phys. Rev. Res.},
  volume = {3},
  issue = {2},
  pages = {023126},
  numpages = {9},
  year = {2021},
  month = {May},
  publisher = {American Physical Society},
  doi = {10.1103/PhysRevResearch.3.023126},
  url = {https://link.aps.org/doi/10.1103/PhysRevResearch.3.023126}
}

@article{11magnon-bistable6,
  title = {Parity-symmetry-breaking quantum phase transition via parametric drive in a cavity magnonic system},
  author = {Zhang, Guo-Qiang and Chen, Zhen and Xiong, Wei and Lam, Chi-Hang and You, J. Q.},
  journal = {Phys. Rev. B},
  volume = {104},
  issue = {6},
  pages = {064423},
  numpages = {10},
  year = {2021},
  month = {Aug},
  publisher = {American Physical Society},
  doi = {10.1103/PhysRevB.104.064423},
  url = {https://link.aps.org/doi/10.1103/PhysRevB.104.064423}
}

@article{12-scattering-magnon-bistable,
  title = {Stochasticity of the magnon parametron},
  author = {Elyasi, Mehrdad and Saitoh, Eiji and Bauer, Gerrit E. W.},
  journal = {Phys. Rev. B},
  volume = {105},
  issue = {5},
  pages = {054403},
  numpages = {12},
  year = {2022},
  month = {Feb},
  publisher = {American Physical Society},
  doi = {10.1103/PhysRevB.105.054403},
  url = {https://link.aps.org/doi/10.1103/PhysRevB.105.054403}
}

@article{1CIMusedtosolvingNPhard1,
  author={Xie, Shanshan and Raman, Siddhartha Raman Sundara and Ni, Can and Wang, Meizhi and Yang, Mengtian and Kulkarni, Jaydeep P.},
  journal={IEEE Journal of Solid-State Circuits}, 
  title={Ising-CIM: A Reconfigurable and Scalable Compute Within Memory Analog Ising Accelerator for Solving Combinatorial Optimization Problems}, 
  year={2022},
  volume={57},
  number={11},
  pages={3453-3465},
  doi={10.1109/JSSC.2022.3176610}
}

@article{1CIMusedtosolvingNPhard2,
  title = {Coherent Ising machine based on degenerate optical parametric oscillators},
  author = {Wang, Zhe and Marandi, Alireza and Wen, Kai and Byer, Robert L. and Yamamoto, Yoshihisa},
  journal = {Phys. Rev. A},
  volume = {88},
  issue = {6},
  pages = {063853},
  numpages = {9},
  year = {2013},
  month = {Dec},
  publisher = {American Physical Society},
  doi = {10.1103/PhysRevA.88.063853},
  url = {https://link.aps.org/doi/10.1103/PhysRevA.88.063853}
}

@article{1CIMusedtosolvingNPhard3,
    author = {Takahiro Inagaki  and Yoshitaka Haribara  and Koji Igarashi  and Tomohiro Sonobe  and Shuhei Tamate  and Toshimori Honjo  and Alireza Marandi  and Peter L. McMahon  and Takeshi Umeki  and Koji Enbutsu  and Osamu Tadanaga  and Hirokazu Takenouchi  and Kazuyuki Aihara  and Ken-ichi Kawarabayashi  and Kyo Inoue  and Shoko Utsunomiya  and Hiroki Takesue },
    title = {A coherent Ising machine for 2000-node optimization problems},
    journal = {Science},
    volume = {354},
    number = {6312},
    pages = {603-606},
    year = {2016},
    doi = {10.1126/science.aah4243},
    URL = {https://www.science.org/doi/abs/10.1126/science.aah4243}
}

@article{10.1038/nphoton.2014.249,
title = {Network of time-multiplexed optical parametric oscillators as a coherent {I}sing machine},
  author = {Marandi, Alireza and Wang, Zhe and Takata, Kenta and Byer, Robert L. and Yamamoto, Yoshihisa},
  journal = {Nature Photonics},
  volume = {8},
  issue = {12},
  pages = {937-942},
  year = {2014},
  month = {Dec.},
  doi = {10.1038/nphoton.2014.249},
  url = {https://doi.org/10.1038/nphoton.2014.249}
}

@article{2CIMusedtogeneratingneuralnetwork1,
    author = {Yamamoto, Y. and Leleu, T. and Ganguli, S. and Mabuchi, H.},
    title = {Coherent Ising machines—Quantum optics and neural network Perspectives},
    journal = {Applied Physics Letters},
    volume = {117},
    number = {16},
    pages = {160501},
    year = {2020},
    month = {10},
    issn = {0003-6951},
    doi = {10.1063/5.0016140},
    url = {https://doi.org/10.1063/5.0016140},
}

@article{2CIMusedtogeneratingneuralnetwork2,
    author = {Bo Lu and Chen-Rui Fan and Lu Liu and Kai Wen and Chuan Wang},
    journal = {Opt. Express},
    number = {3},
    pages = {3676--3684},
    publisher = {Optica Publishing Group},
    title = {Speed-up coherent Ising machine with a spiking neural network},
    volume = {31},
    month = {Jan},
    year = {2023},
    url = {https://opg.optica.org/oe/abstract.cfm?URI=oe-31-3-3676},
    doi = {10.1364/OE.479903},
}

@article{10.1126/science.aah5178,
author = {Peter L. McMahon  and Alireza Marandi  and Yoshitaka Haribara  and Ryan Hamerly  and Carsten Langrock  and Shuhei Tamate  and Takahiro Inagaki  and Hiroki Takesue  and Shoko Utsunomiya  and Kazuyuki Aihara  and Robert L. Byer  and M. M. Fejer  and Hideo Mabuchi  and Yoshihisa Yamamoto },
title = {A fully programmable 100-spin coherent {I}sing machine with all-to-all connections},
journal = {Science},
volume = {354},
number = {6312},
pages = {614-617},
year = {2016},
doi = {10.1126/science.aah5178},
URL = {https://www.science.org/doi/abs/10.1126/science.aah5178}
}

@article{3CIMusedtomakinglogiccircuits1,
    author = {Toshimori Honjo  and Tomohiro Sonobe  and Kensuke Inaba  and Takahiro Inagaki  and Takuya Ikuta  and Yasuhiro Yamada  and Takushi Kazama  and Koji Enbutsu  and Takeshi Umeki  and Ryoichi Kasahara  and Ken-ichi Kawarabayashi  and Hiroki Takesue },
    title = {100,000-spin coherent Ising machine},
    journal = {Science Advances},
    volume = {7},
    number = {40},
    pages = {eabh0952},
    year = {2021},
    doi = {10.1126/sciadv.abh0952},
    URL = {https://www.science.org/doi/abs/10.1126/sciadv.abh0952}
}

@ARTICLE{3CIMusedtomakinglogiccircuits2,
  author={Aonishi, Toru and Nagasawa, Tatsuya and Koizumi, Toshiyuki and Gunathilaka, Mastiyage Don Sudeera Hasaranga and Mimura, Kazushi and Okada, Masato and Kako, Satoshi and Yamamoto, Yoshihisa},
  journal={IEEE Access}, 
  title={Highly Versatile FPGA-Implemented Cyber Coherent Ising Machine}, 
  year={2024},
  volume={12},
  number={},
  pages={175843-175865},
  doi={10.1109/ACCESS.2024.3504008}
}

@article{3CIMusedtomakinglogiccircuits3,
  author  = {Mohseni, Naeimeh and McMahon, Peter L. and Byrnes, Tim},
  title   = {Ising machines as hardware solvers of combinatorial optimization problems},
  journal = {Nature Reviews Physics},
  volume  = {4},
  number  = {6},
  pages   = {363--379},
  year    = {2022},
  month   = {5},
  issn    = {2522-5820},
  doi     = {10.1038/s42254-022-00440-8},
  url     = {https://doi.org/10.1038/s42254-022-00440-8},
}

@article{4CIMinexperiment1,
    author = {Ryan Hamerly  and Takahiro Inagaki  and Peter L. McMahon  and Davide Venturelli  and Alireza Marandi  and Tatsuhiro Onodera  and Edwin Ng  and Carsten Langrock  and Kensuke Inaba  and Toshimori Honjo  and Koji Enbutsu  and Takeshi Umeki  and Ryoichi Kasahara  and Shoko Utsunomiya  and Satoshi Kako  and Ken-ichi Kawarabayashi  and Robert L. Byer  and Martin M. Fejer  and Hideo Mabuchi  and Dirk Englund  and Eleanor Rieffel  and Hiroki Takesue  and Yoshihisa Yamamoto },
    title = {Experimental investigation of performance differences between coherent Ising machines and a quantum annealer},
    journal = {Science Advances},
    volume = {5},
    number = {5},
    pages = {eaau0823},
    year = {2019},
    doi = {10.1126/sciadv.aau0823},
    URL = {https://www.science.org/doi/abs/10.1126/sciadv.aau0823}
}

@article{4CIMinexperiment2,
    title = {Efficient sampling of ground and low-energy Ising spin configurations with a coherent Ising machine},
    author = {Ng, Edwin and Onodera, Tatsuhiro and Kako, Satoshi and McMahon, Peter L. and Mabuchi, Hideo and Yamamoto, Yoshihisa},
    journal = {Phys. Rev. Res.},
    volume = {4},
    issue = {1},
    pages = {013009},
    numpages = {22},
    year = {2022},
    month = {Jan},
    publisher = {American Physical Society},
    doi = {10.1103/PhysRevResearch.4.013009},
    url = {https://link.aps.org/doi/10.1103/PhysRevResearch.4.013009}
}

@article{4CIMinexperiment3,
  title = {Observing a Phase Transition in a Coherent Ising Machine},
  author = {Takesue, Hiroki and Yamada, Yasuhiro and Inaba, Kensuke and Ikuta, Takuya and Yonezu, Yuya and Inagaki, Takahiro and Honjo, Toshimori and Kazama, Takushi and Enbutsu, Koji and Umeki, Takeshi and Kasahara, Ryoichi},
  journal = {Phys. Rev. Appl.},
  volume = {19},
  issue = {3},
  pages = {L031001},
  numpages = {6},
  year = {2023},
  month = {Mar},
  publisher = {American Physical Society},
  doi = {10.1103/PhysRevApplied.19.L031001},
  url = {https://link.aps.org/doi/10.1103/PhysRevApplied.19.L031001}
}

@article{5CIMbasedon3,
  title = {All-Optical Scalable Spatial Coherent Ising Machine},
  author = {Calvanese Strinati, Marcello and Pierangeli, Davide and Conti, Claudio},
  journal = {Phys. Rev. Appl.},
  volume = {16},
  issue = {5},
  pages = {054022},
  numpages = {7},
  year = {2021},
  month = {Nov},
  publisher = {American Physical Society},
  doi = {10.1103/PhysRevApplied.16.054022},
  url = {https://link.aps.org/doi/10.1103/PhysRevApplied.16.054022}
}

@article{5CIMbasedon5,
    author = {Ezawa, Motohiko and Lebrasseur , Eric and Mita , Yoshio},
    title = {Ising Machine Based on Bistable Microelectromechanical Systems},
    journal = {Journal of the Physical Society of Japan},
    volume = {91},
    number = {11},
    pages = {114601},
    year = {2022},
    doi = {10.7566/JPSJ.91.114601},
    URL = {https://doi.org/10.7566/JPSJ.91.114601}
}

@article{1Multi-enhanced-coupling2,
  title = {Quantum many-body scars with unconventional superconducting pairing symmetries via multibody interactions},
  author = {Imai, Shohei and Tsuji, Naoto},
  journal = {Phys. Rev. Res.},
  volume = {7},
  issue = {1},
  pages = {013064},
  numpages = {18},
  year = {2025},
  month = {Jan},
  publisher = {American Physical Society},
  doi = {10.1103/PhysRevResearch.7.013064},
  url = {https://link.aps.org/doi/10.1103/PhysRevResearch.7.013064}
}

@article{1Multi-enhanced-coupling3,
  title = {Strong long-range spin-spin coupling via a Kerr magnon interface},
  author = {Xiong, Wei and Tian, Miao and Zhang, Guo-Qiang and You, J. Q.},
  journal = {Phys. Rev. B},
  volume = {105},
  issue = {24},
  pages = {245310},
  numpages = {7},
  year = {2022},
  month = {Jun},
  publisher = {American Physical Society},
  doi = {10.1103/PhysRevB.105.245310},
  url = {https://link.aps.org/doi/10.1103/PhysRevB.105.245310}
}

@article{2Multi-dissipative-coupling1,
  title = {Strong and noise-tolerant entanglement in dissipative optomechanics},
  author = {Chen, Jiaojiao and Xiong, Wei and Wang, Dong and Ye, Liu},
  journal = {Phys. Rev. A},
  volume = {111},
  issue = {5},
  pages = {053512},
  numpages = {7},
  year = {2025},
  month = {May},
  publisher = {American Physical Society},
  doi = {10.1103/PhysRevA.111.053512},
  url = {https://link.aps.org/doi/10.1103/PhysRevA.111.053512}
}

@article{2Multi-dissipative-coupling2,
  author  = {Wang, Yimin and Xiong, Wei and Xu, Zhiyong and Zhang, Guo-Qiang and You, Jian-Qiang},
  title   = {Dissipation-induced nonreciprocal magnon blockade in a magnon-based hybrid system},
  journal = {Science China Physics, Mechanics \& Astronomy},
  volume  = {65},
  number  = {6},
  pages   = {260314},
  year    = {2022},
  month   = {4},
  issn    = {1869-1927},
  doi     = {10.1007/s11433-021-1880-7},
  url     = {https://doi.org/10.1007/s11433-021-1880-7},
}

@article{3photons-phonons-and-magnons-1,
  title = {Coherent Coupling between Phonons, Magnons, and Photons},
  author = {Shen, Zhen and Xu, Guan-Ting and Zhang, Mai and Zhang, Yan-Lei and Wang, Yu and Chai, Cheng-Zhe and Zou, Chang-Ling and Guo, Guang-Can and Dong, Chun-Hua},
  journal = {Phys. Rev. Lett.},
  volume = {129},
  issue = {24},
  pages = {243601},
  numpages = {7},
  year = {2022},
  month = {Dec},
  publisher = {American Physical Society},
  doi = {10.1103/PhysRevLett.129.243601},
  url = {https://link.aps.org/doi/10.1103/PhysRevLett.129.243601}
}

@article{3photons-phonons-and-magnons-2,
  author  = {Amazioug, Mohamed and Teklu, Berihu and Asjad, Muhammad},
  title   = {Enhancement of magnon–photon–phonon entanglement in a cavity magnomechanics with coherent feedback loop},
  journal = {Scientific Reports},
  volume  = {13},
  number  = {1},
  pages   = {3833},
  year    = {2023},
  month   = mar,
  issn    = {2045-2322},
  doi     = {10.1038/s41598-023-30693-x},
  url     = {https://doi.org/10.1038/s41598-023-30693-x},
}

@article{3photons-phonons-and-magnons-3,
  title = {Magnon-Photon-Phonon Entanglement in Cavity Magnomechanics},
  author = {Li, Jie and Zhu, Shi-Yao and Agarwal, G. S.},
  journal = {Phys. Rev. Lett.},
  volume = {121},
  issue = {20},
  pages = {203601},
  numpages = {6},
  year = {2018},
  month = {Nov},
  publisher = {American Physical Society},
  doi = {10.1103/PhysRevLett.121.203601},
  url = {https://link.aps.org/doi/10.1103/PhysRevLett.121.203601}
}

@article{3photons-phonons-and-magnons-4,
  title = {Magnon-assisted photon-phonon conversion in the presence of structured environments},
  author = {Qi, Shi-fan and Jing, Jun},
  journal = {Phys. Rev. A},
  volume = {103},
  issue = {4},
  pages = {043704},
  numpages = {10},
  year = {2021},
  month = {Apr},
  publisher = {American Physical Society},
  doi = {10.1103/PhysRevA.103.043704},
  url = {https://link.aps.org/doi/10.1103/PhysRevA.103.043704}
}

@misc{3photons-phonons-and-magnons-5,
      title={Tunable Entanglement in Cavity-Magnon Optomechanics}, 
      author={Ming-Yue Liu and Xian-Xian Huang and Wei Xiong},
      year={2024},
      eprint={2404.15111},
      archivePrefix={arXiv}
}

@article{4triple-squeeze-1,
  title = {All-optical polarization-state engineering in quantum cavity optomagnonics},
  author = {Liang, Zhu and Li, Jiahua and Wu, Ying},
  journal = {Phys. Rev. A},
  volume = {107},
  issue = {3},
  pages = {033701},
  numpages = {21},
  year = {2023},
  month = {Mar},
  publisher = {American Physical Society},
  doi = {10.1103/PhysRevA.107.033701},
  url = {https://link.aps.org/doi/10.1103/PhysRevA.107.033701}
}

@misc{4triple-squeeze-2,
      title={Atom-Molecule Superradiance and Entanglement with Cavity-Mediated Three-Body Interactions}, 
      author={Yun Chen and Yuqi Wang and Jingjun You and Yingqi Liu and Su Yi and Yuangang Deng},
      year={2025},
      eprint={2501.09497},
      archivePrefix={arXiv}
}

@article{5triple-feedback-control1,
  title = {Quantum Fokker-Planck Master Equation for Continuous Feedback Control},
  author = {Annby-Andersson, Bj\"orn and Bakhshinezhad, Faraj and Bhattacharyya, Debankur and De Sousa, Guilherme and Jarzynski, Christopher and Samuelsson, Peter and Potts, Patrick P.},
  journal = {Phys. Rev. Lett.},
  volume = {129},
  issue = {5},
  pages = {050401},
  numpages = {8},
  year = {2022},
  month = {Jul},
  publisher = {American Physical Society},
  doi = {10.1103/PhysRevLett.129.050401},
  url = {https://link.aps.org/doi/10.1103/PhysRevLett.129.050401}
}

@article{5triple-feedback-control2,
  title={Feedback control of quantum correlations in a cavity magnomechanical system with magnon squeezing},
  author={Amazioug, Mohamed and Singh, Shailendra and Teklu, Berihu and Asjad, Muhammad},
  journal={Entropy},
  volume={25},
  number={10},
  pages={1462},
  year={2023},
  month = {Oct},
  publisher={MDPI},
  doi = {https://doi.org/10.3390/e25101462}
}

@article{6partite-control-partite1,
  title = {Driven dissipative quantum dynamics in a cavity magnon-polariton system},
  author = {Zhao, Guogan and Wang, Yong and Qian, X.-F.},
  journal = {Phys. Rev. B},
  volume = {104},
  issue = {13},
  pages = {134423},
  numpages = {10},
  year = {2021},
  month = {Oct},
  publisher = {American Physical Society},
  doi = {10.1103/PhysRevB.104.134423},
  url = {https://link.aps.org/doi/10.1103/PhysRevB.104.134423}
}

@article{6partite-control-partite2,
  title = {Dissipation-induced magnon-photon entanglement in a squeezed vacuum reservoir},
  author = {Xiao, Yang and Xia, K.},
  journal = {Phys. Rev. B},
  volume = {111},
  issue = {6},
  pages = {064414},
  numpages = {8},
  year = {2025},
  month = {Feb},
  publisher = {American Physical Society},
  doi = {10.1103/PhysRevB.111.064414},
  url = {https://link.aps.org/doi/10.1103/PhysRevB.111.064414}
}

@article{6partite-control-partite3,
  title = {Switchable superradiant phase transition with Kerr magnons},
  author = {Liu, Gang and Xiong, Wei and Ying, Zu-Jian},
  journal = {Phys. Rev. A},
  volume = {108},
  issue = {3},
  pages = {033704},
  numpages = {7},
  year = {2023},
  month = {Sep},
  publisher = {American Physical Society},
  doi = {10.1103/PhysRevA.108.033704},
  url = {https://link.aps.org/doi/10.1103/PhysRevA.108.033704}
}

@article{7dissipation-with-stable-states1,
  title = {Generation of long-lived $W$ states via reservoir engineering in dissipatively coupled systems},
  author = {Zhang, Guo-Qiang and Feng, Wei and Xiong, Wei and Su, Qi-Ping and Yang, Chui-Ping},
  journal = {Phys. Rev. A},
  volume = {107},
  issue = {1},
  pages = {012410},
  numpages = {10},
  year = {2023},
  month = {Jan},
  publisher = {American Physical Society},
  doi = {10.1103/PhysRevA.107.012410},
  url = {https://link.aps.org/doi/10.1103/PhysRevA.107.012410}
}

@article{7dissipation-with-stable-states2,
  title = {Coherent Quantum Dynamics in Steady-State Manifolds of Strongly Dissipative Systems},
  author = {Zanardi, Paolo and Campos Venuti, Lorenzo},
  journal = {Phys. Rev. Lett.},
  volume = {113},
  issue = {24},
  pages = {240406},
  numpages = {5},
  year = {2014},
  month = {Dec},
  publisher = {American Physical Society},
  doi = {10.1103/PhysRevLett.113.240406},
  url = {https://link.aps.org/doi/10.1103/PhysRevLett.113.240406}
}

@article{7dissipation-with-stable-states3,
  title = {Stabilizing effect of driving and dissipation on quantum metastable states},
  author = {Valenti, Davide and Carollo, Angelo and Spagnolo, Bernardo},
  journal = {Phys. Rev. A},
  volume = {97},
  issue = {4},
  pages = {042109},
  numpages = {7},
  year = {2018},
  month = {Apr},
  publisher = {American Physical Society},
  doi = {10.1103/PhysRevA.97.042109},
  url = {https://link.aps.org/doi/10.1103/PhysRevA.97.042109}
}

@article{7dissipation-with-stable-states4,
  title = {Quantum dynamics of a two-state system in a dissipative environment},
  author = {Ao, Ping and Rammer, Jo/rgen},
  journal = {Phys. Rev. B},
  volume = {43},
  issue = {7},
  pages = {5397--5418},
  numpages = {0},
  year = {1991},
  month = {Mar},
  publisher = {American Physical Society},
  doi = {10.1103/PhysRevB.43.5397},
  url = {https://link.aps.org/doi/10.1103/PhysRevB.43.5397}
}

@article{self-kerr,
  title = {Mechanical Bistability in Kerr-modified Cavity Magnomechanics},
  author = {Shen, Rui-Chang and Li, Jie and Fan, Zhi-Yuan and Wang, Yi-Pu and You, J. Q.},
  journal = {Phys. Rev. Lett.},
  volume = {129},
  issue = {12},
  pages = {123601},
  numpages = {7},
  year = {2022},
  month = {Sep},
  publisher = {American Physical Society},
  doi = {10.1103/PhysRevLett.129.123601},
  url = {https://link.aps.org/doi/10.1103/PhysRevLett.129.123601}
}

@article{santiouhe,
  title = {Enhanced Tripartite Interactions in Spin-Magnon-Mechanical Hybrid Systems},
  author = {Hei, Xin-Lei and Li, Peng-Bo and Pan, Xue-Feng and Nori, Franco},
  journal = {Phys. Rev. Lett.},
  volume = {130},
  issue = {7},
  pages = {073602},
  numpages = {10},
  year = {2023},
  month = {Feb},
  publisher = {American Physical Society},
  doi = {10.1103/PhysRevLett.130.073602},
  url = {https://link.aps.org/doi/10.1103/PhysRevLett.130.073602}
}

@article{squeeze--pump,
  title = {Optomechanical-interface-induced strong spin-magnon coupling},
  author = {Xiong, Wei and Wang, Mingfeng and Zhang, Guo-Qiang and Chen, Jiaojiao},
  journal = {Phys. Rev. A},
  volume = {107},
  issue = {3},
  pages = {033516},
  numpages = {12},
  year = {2023},
  month = {Mar},
  publisher = {American Physical Society},
  doi = {10.1103/PhysRevA.107.033516},
  url = {https://link.aps.org/doi/10.1103/PhysRevA.107.033516}
}

@article{POSI2NEGA,
    author = {Fan, Xiao-Hong and Zhang, Yi-Ning and Yu, Jun-Po and Liu, Ming-Yue and He, Wen-Di and Li, Hai-Chao and Xiong, Wei},
    title = {Nonreciprocal Unconventional Photon Blockade with Kerr Magnons},
    journal = {Advanced Quantum Technologies},
    volume = {7},
    number = {8},
    pages = {2400043},
    doi = {https://doi.org/10.1002/qute.202400043},
    url = {https://advanced.onlinelibrary.wiley.com/doi/abs/10.1002/qute.202400043},
    year = {2024}
}

@article{10.1021/acs.jctc.3c00943,
  title = {Encoding Molecular Docking for Quantum Computers},
  author = {Zha, Jinyin and Su, Jiaqi and Li, Tiange and Cao, Chongyu and Ma, Yin and Wei, Hai and Huang, Zhiguo and Qian, Ling and Wen, Kai and Zhang, Jian},
  journal = {J. Chem. Theory Comput.},
  volume = {19},
  issue = {24},
  pages = {9018},
  numpages = {6},
  year = {2023},
  month = {Dec},
  publisher = {American Chemical Society},
  doi = {10.1021/acs.jctc.3c00943},
  url = { https://doi.org/10.1021/acs.jctc.3c00943}
}

@article{10.1007/s11433-023-2147-3,
  title = {Optical experimental solution for the multiway number partitioning problem and its application to computing power scheduling},
  author = {Wen, Jingwei and Wang, Zhenming and Huang, Zhiguo and Cai, Dunbo and Jia, Bingjie and Cao, Chongyu and Ma, Yin and Wei, Hai and Wen, Kai and Qian, Ling},
  journal = {Science China Physics, Mechanics {\&} Astronomy},
  volume = {66},
  issue = {9},
  pages = {290313},
  year = {2023},
  month = {Aug},
  doi = {10.1007/s11433-023-2147-3},
  url = { https://doi.org/10.1007/s11433-023-2147-3}
}

@article{10.1103/c91r-8t3h,
  title = {Parametric instability landscape of coupled Kerr parametric oscillators},
  author = {Ameye, Orjan and Eichler, Alexander and Zilberberg, Oded},
  journal = {Phys. Rev. Res.},
  volume = {7},
  issue = {3},
  pages = {033204},
  numpages = {16},
  year = {2025},
  month = {Aug},
  publisher = {American Physical Society},
  doi = {10.1103/c91r-8t3h},
  url = {https://link.aps.org/doi/10.1103/c91r-8t3h}
}

@article{10.1126/science.aaf2941,
author = {Chen Wang  and Yvonne Y. Gao  and Philip Reinhold  and R. W. Heeres  and Nissim Ofek  and Kevin Chou  and Christopher Axline  and Matthew Reagor  and Jacob Blumoff  and K. M. Sliwa  and L. Frunzio  and S. M. Girvin  and Liang Jiang  and M. Mirrahimi  and M. H. Devoret  and R. J. Schoelkopf },
title = {A Schrödinger cat living in two boxes},
journal = {Science},
volume = {352},
number = {6289},
pages = {1087-1091},
year = {2016},
doi = {10.1126/science.aaf2941},
URL = {https://www.science.org/doi/abs/10.1126/science.aaf2941}
}

@article{10.1016/j.cpc.2012.02.021,
      title = {QuTiP: An open-source Python framework for the dynamics of open quantum systems},
      journal = {Computer Physics Communications},
      volume = {183},
      number = {8},
      pages = {1760-1772},
      year = {2012},
      issn = {0010-4655},
      doi = {https://doi.org/10.1016/j.cpc.2012.02.021},
      url = {https://www.sciencedirect.com/science/article/pii/S0010465512000835},
      author = {J.R. Johansson and P.D. Nation and Franco Nori}
}

@article{10.1016/j.cpc.2012.11.019,
      title = {QuTiP 2: A Python framework for the dynamics of open quantum systems},
      journal = {Computer Physics Communications},
      volume = {184},
      number = {4},
      pages = {1234-1240},
      year = {2013},
      issn = {0010-4655},
      doi = {https://doi.org/10.1016/j.cpc.2012.11.019},
      url = {https://www.sciencedirect.com/science/article/pii/S0010465512003955},
      author = {J.R. Johansson and P.D. Nation and Franco Nori}
}

@article{2412.04705,
      title={QuTiP 5: The Quantum Toolbox in Python}, 
      author={Neill Lambert and Eric Giguère and Paul Menczel and Boxi Li and Patrick Hopf and Gerardo Suárez and Marc Gali and Jake Lishman and Rushiraj Gadhvi and Rochisha Agarwal and Asier Galicia and Nathan Shammah and Paul Nation and J. R. Johansson and Shahnawaz Ahmed and Simon Cross and Alexander Pitchford and Franco Nori},
      journal = {arXiv:2412.04705},
      year={2024},
      url={https://arxiv.org/abs/2412.04705}, 
}
\end{document}